\documentclass[a4paper,twocolumn,aps,prl,floatfix,reprint,superscriptaddress,showpacs]{revtex4-1}
\usepackage[latin9]{inputenc}
\usepackage{color}
\usepackage{mathrsfs}
\usepackage{amsmath}
\usepackage{amssymb}
\usepackage{graphicx}
\usepackage[bookmarks=false,
 breaklinks=false,pdfborder={0 0 1},backref=false,colorlinks=true]
 {hyperref}
\hypersetup{
 letterpaper=true,pdfstartview=FitV,linkcolor=blue,citecolor=blue,urlcolor=blue}

\makeatletter

\pdfpageheight\paperheight
\pdfpagewidth\paperwidth

\pdfoutput=1
\usepackage{mathrsfs}
\usepackage{color}
\usepackage{bm}
\usepackage{xspace}
\usepackage{epstopdf}
\usepackage{dcolumn}
\usepackage{longtable}
\usepackage{amsmath}    
\allowdisplaybreaks[4]  

\makeatother

\begin{document}
\title{Linear and nonlinear optical responses in Green\textquoteright s function
formula }
\date{\today}
\author{Maoyuan Wang}
\affiliation{Department of Physics, Xiamen University, Xiamen, China. }
\author{Jianhui Zhou}
\email{jhzhou@hmfl.ac.cn}

\affiliation{Anhui Provincial Key Laboratory of Low-Energy Quantum Materials and
Devices, High Magnetic Field Laboratory, HFIPS, Chinese Academy of
Sciences, Hefei, Anhui 230031, China}
\author{Yugui Yao}
\email{ygyao@bit.edu.cn}

\affiliation{Key Lab of advanced optoelectronic quantum architecture and measurement
(MOE), Beijing Key Lab of Nanophotonics \& Ultrafine Optoelectronic
Systems and School of Physics, Beijing Institute of Technology, Beijing
100081, China}
\begin{abstract}
Linear and nonlinear optical effect has been widely discussed in large
quantity of materials using theoretical or experimental methods. Except
linear optical conductivity, higher-order nonlinear responses are
not studied fully. Starting from density operator method, we derive
optical conductivities of different orders in Green\textquoteright s
function formula, and also connect them to novel physical quantities,
such as Berry curvature, Berry curvature dipole, third-order nonlinear
Hall conductivity and so on. Based on the advantages of Green\textquoteright s
function formulas, we believe that these formulas have a lot of benefits
for many-body effect study in high-order nonlinear optical responses.
\end{abstract}
\maketitle

\section{Introduction}

Nonlinear optical is one of the most important topic in condensed
matter physics\citep{boyd2003nonlinear}. Except the well studied
linear optical and magneto-optical response, the nonlinear response
involving multi-photon process provide more rich physics, which are
both theoretically interested and exhibiting potentially applications.
Until now, numbers of nonlinear effects has been put forward. For
instance the second-order-harmonic generation (SHG)\citep{SHG,N2019-Wushiwei-giantSHG,prb2000-sipe-2nd-length-gauge-formula}
related with nonlinear magneto-optical effect\citep{review-mo-ebert1996magneto,review-mo-kirilyuk2002nonlinear,review-mo-pustogowa1996theory},
can be applied for magnetism detection\citep{SHG-application}. Another
important class of high order response is optical reflection(OR)\citep{boyd2003nonlinear}.
In second order, OR include shift current\citep{y1992-BVPE,prb2000-sipe-2nd-length-gauge-formula,prl2000-BVPE,am2010-BVPE,npj2016-BVPE,nc2017-BVPE,prl2017-BVPE}
and injection current\citep{prb2000-sipe-2nd-length-gauge-formula,y2010-CPGE,prb2011-CPGE,nnano2012-CPGE,nnano2014-CPGE},
which is responsible for bulk photovoltaic effect (BPVE) and circular
photogalvanic effect (CPGE) respectively. Due to the generated dc
current in bulk materials, the BPVE can be applied in solar cell\citep{prb1981-BVPE-formula,prl-Young-BVPE-photovoltaic,N2013-photovoltaic,S2015-solar-cells-1,S2015-solar-cells-2,S2015-solar-cells-3,nc2013-photovoltaic}. 

Historically, the nonlinear susceptibilities are in terms of matrix
elements and bands energy derived in fully quantum treatment with
light matter interaction represented by either so called length gauge
or velocity gauge\citep{prb1979-BVPE-formula,prb1981-BVPE-formula,prb1993-sipe-formula,prb1995-sipe-formula-length-gauge,prb1996-sipe-formula-2nd,prb2000-sipe-2nd-length-gauge-formula,prb2019-JMoore-diagrammatic-formula}.
Recently, along with the theoretical development of geometry phase
in electronic bands\citep{RMP-Niu-BerryPhaseInSolid}, the high order
susceptibilities has been modified into berry curvature related quantities\citep{prb-JMoore-OpticalGyrotropy,prl-JMoore-Gyrotropy-axion,prl-JMoore-semiclassical-CPGE-LPGE,nc-JMoore-QuantizedCPGE,prl-FuLiang-NLHE,sciadv-morimoto-floquet,arxiv-deyo2009-semiclassical,arxiv-YanBH-bcdipole-QuantumMatrix,fop2019-GaoYang-semiclassical},
and significant photocurrent was observed in topological materials
such as Weyl semimetals\citep{nm2019-exp-BVPE,np2017-exp-TaAs-CPGE,np-XuSY-interband-bcDipole-exp,nc2019-exp-WTe2-sc,N-MaQiong-NLHE}.
The berry curvature dipole\citep{prl-FuLiang-NLHE} was proposed from
semiclassical framework, which can generate transverse current by
applying an longitudinal electric field in inversion breaking materials\citep{prl-FuLiang-NLHE,N-MaQiong-NLHE}
suggesting a nonlinear Hall effect(NLHE). It was shown that the susceptibilities
for injection current can be modified in terms of inter-bands Berry
curvature dipole\citep{np-XuSY-interband-bcDipole-exp}. Also, combining
with Keldysh formula and Floquet treatment for light matter interaction,
it have derived numbers of high order susceptibilities in terms of
geometric quantities\citep{sciadv-morimoto-floquet}. 

To date, most of these theory are derived in Lehmann representation,
and only few of them is in Green's function formulas leading them
hard to consider many-body effect. It is well known that in anomalous
Hall effect, the disorder effect is not neglected in large range of
materials\citep{RMP-AHE-nagaosa,potp2006-nagaosa-nonequilibirum,prb2008-nagaosa-nonequilibirum}.
For NLHE, there have taken into account disorder effect in Boltzmann's
framework\citep{nc2019-LuHZ-disorder-NLHE}, which might not be systematic
and complete enough. The many-body formulation for nonlinear optical
within the fully quantum treatment need further theoretical efforts. 

In this paper, the derivation of optical conductivity of linear and
nonlinear in Green\textquoteright s function formula are presented.
At the beginning, we list the general method of density operator to
deal with linear or nonlinear response cases, and classify response
functions according to their orders. Then we consider the case of
optical response, and derive linear and nonlinear optical conductivities
in Green\textquoteright s function formula up to third-order. We also
list some novel physical quantities in Green\textquoteright s function
formula, such as berry curvature, berry curvature dipole, shift current,
third-order nonlinear hall conductivity and so on, from linear combination
of linear or nonlinear optical conductivities. Furthermore, taking
the advantages of Green\textquoteright s function formulas, we discuss
many-body effect and take disorder effect as an example. 

\section{General linear or nonlinear response}

\subsection{Density operator under perturbation}

For a system $H_{0}$ with a perturbation , the total Hamiltonian
\begin{equation}
H=H_{0}+V_{1}(t)+V_{2}(t)+\cdots
\end{equation}
, where n in $V_{n}$ marked the n-th order of perturbation. And the
equation of motion for density operator is $i\hbar\dot{\rho}\left(t\right)=\left[H,\rho\right]$,
where $\rho_{0}$ is the unperturbated density operator. According
to standard perturbation procedure, we replace $\underset{n}{\sum}V_{n}(t)$
and $\rho$ with $\underset{n}{\sum}\lambda^{n}V_{n}(t)$ and $\underset{n}{\sum}\lambda^{n}\rho_{n}(t)$
in the equation of motion, and separate terms according to the order
of $\lambda$, then we can get the equation of motion for different
order of density operator:

\begin{eqnarray}
i\hbar\dot{\rho}_{0} & = & \left[H_{0},\rho_{0}\right]\\
i\hbar\dot{\rho}_{1}\left(t\right) & = & \left[H_{0},\rho_{1}\left(t\right)\right]+\left[V_{1}\left(t\right),\rho_{0}\right]\\
i\hbar\dot{\rho}_{2}\left(t\right) & = & \left[H_{0},\rho_{2}\left(t\right)\right]+\left[V_{1}\left(t\right),\rho_{1}\left(t\right)\right]+\left[V_{2}\left(t\right),\rho_{0}\right]\\
i\hbar\dot{\rho}_{3}\left(t\right) & = & \left[H_{0},\rho_{3}\left(t\right)\right]+\left[V_{1}\left(t\right),\rho_{2}\left(t\right)\right]\nonumber \\
 &  & +\left[V_{2}\left(t\right),\rho_{1}\left(t\right)\right]+\left[V_{3}\left(t\right),\rho_{0}\right]\\
 & \vdots\nonumber 
\end{eqnarray}

Since the density operator here is in Schr�dinger picture, it is convenient
to utilize the interaction picture, $\rho^{I}\left(t\right)=e^{iH_{0}t}\rho(t)e^{-iH_{0}t}$.
The equation of motion for density operator can be simplified as:

\begin{eqnarray}
i\hbar\dot{\rho}_{0}^{I} & = & 0\\
i\hbar\dot{\rho}_{1}^{I}\left(t\right) & = & \left[V_{1}^{I}\left(t\right),\rho_{0}^{I}\left(t\right)\right]\\
i\hbar\dot{\rho}_{2}^{I}\left(t\right) & = & \left[V_{2}^{I}\left(t\right),\rho_{0}^{I}\left(t\right)\right]+\left[V_{1}^{I}\left(t\right),\rho_{1}^{I}\left(t\right)\right]\\
i\hbar\dot{\rho}_{3}^{I}\left(t\right) & = & \left[V_{3}^{I}\left(t\right),\rho_{0}^{I}\left(t\right)\right]+\left[V_{2}^{I}\left(t\right),\rho_{1}^{I}\left(t\right)\right]\\
 &  & +\left[V_{1}^{I}\left(t\right),\rho_{2}^{I}\left(t\right)\right]\nonumber \\
 & \vdots\nonumber 
\end{eqnarray}

After integral, the density operator can be simplified as:

\begin{eqnarray}
\rho_{1}^{I}\left(t\right) & = & -\frac{i}{\hbar}\int_{-\infty}^{t}dt^{\prime}\,\left[V_{1}^{I}\left(t^{\prime}\right),\rho_{0}^{I}\right]\\
\rho_{2}^{I}\left(t\right) & = & -\frac{i}{\hbar}\int_{-\infty}^{t}dt^{\prime}\,\left[V_{2}^{I}\left(t^{\prime}\right),\rho_{0}^{I}\right]\\
 &  & +(-\frac{i}{\hbar})^{2}\int_{-\infty}^{t}dt^{\prime}\int_{-\infty}^{t^{\prime}}dt^{\prime\prime}\left[V_{1}^{I}\left(t^{\prime}\right),\left[V_{1}^{I}\left(t^{\prime\prime}\right),\rho_{0}^{I}\right]\right]\nonumber \\
\rho_{3}^{I}\left(t\right) & = & -\frac{i}{\hbar}\int_{-\infty}^{t}dt^{\prime}\,\left[V_{3}^{I}\left(t^{\prime}\right),\rho_{0}^{I}\right]\\
 &  & +(-\frac{i}{\hbar})^{2}\int_{-\infty}^{t}dt^{\prime}\int_{-\infty}^{t^{\prime}}dt^{\prime\prime}\left[V_{2}^{I}\left(t^{\prime}\right),\left[V_{1}^{I}\left(t^{\prime\prime}\right),\rho_{0}^{I}\right]\right]\nonumber \\
 &  & +(-\frac{i}{\hbar})^{2}\int_{-\infty}^{t}dt^{\prime}\int_{-\infty}^{t^{\prime}}dt^{\prime\prime}\left[V_{1}^{I}\left(t^{\prime}\right),\left[V_{2}^{I}\left(t^{\prime\prime}\right),\rho_{0}^{I}\right]\right]\nonumber \\
 &  & +(-\frac{i}{\hbar})^{3}\int_{-\infty}^{t}dt^{\prime}\int_{-\infty}^{t^{\prime}}dt^{\prime\prime}\int_{-\infty}^{t^{\prime\prime}}dt^{\prime\prime\prime}\nonumber \\
 &  & \,\,\,\,\,\,\,\,\,\,\,\,\,\,\,\,\,\,\,\,\left[V_{1}^{I}\left(t^{\prime}\right),\left[V_{1}^{I}\left(t^{\prime\prime}\right),\left[V_{1}^{I}\left(t^{\prime\prime\prime}\right),\rho_{0}^{I}\right]\right]\right]\nonumber \\
 & \vdots\nonumber 
\end{eqnarray}

Here, $\rho_{0}(t)=\rho_{0}^{I}(t)=\rho_{0}$. For n-th order density
operator $\rho_{n}^{I}\left(t\right)$, it has $2^{n-1}$terms.

\begin{figure}[t]
\includegraphics[width=1\columnwidth]{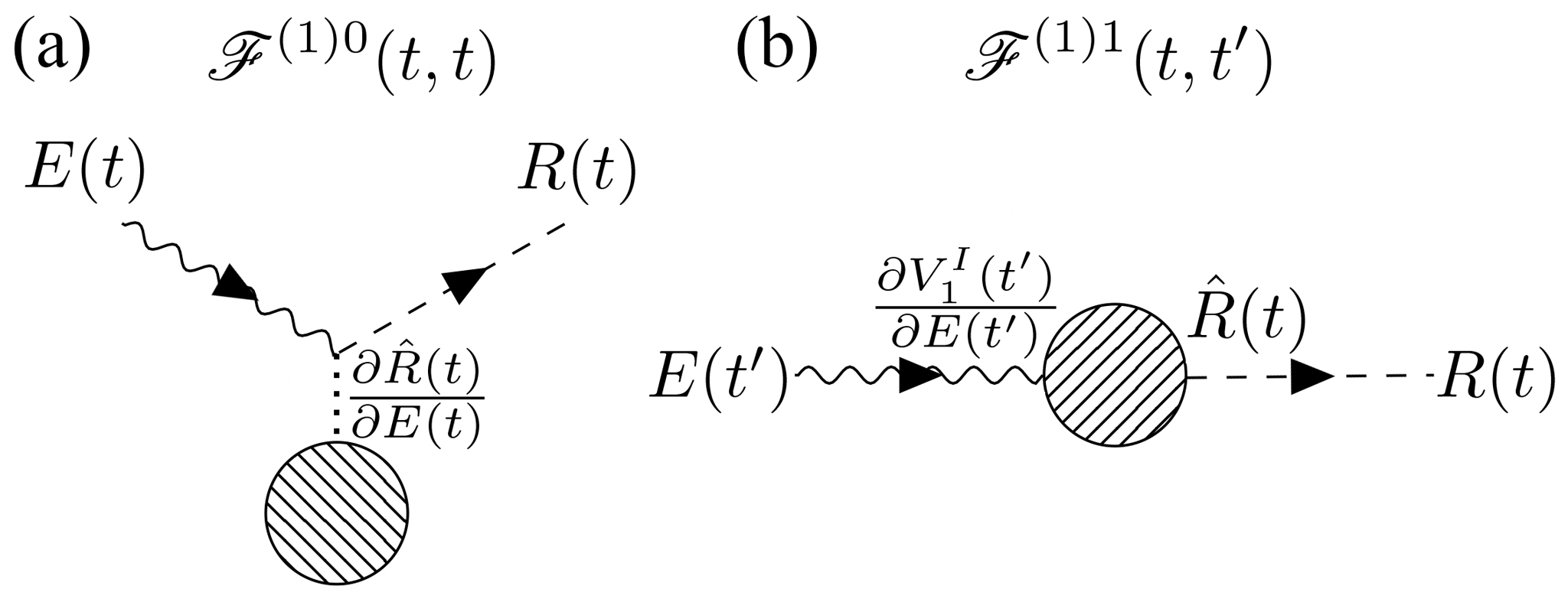} \caption{\textbf{First-order response Feynman diagrams} (a) $\mathscr{F}^{\left(1\right)0}(t,t)$
and (b) $\mathscr{F}^{\left(1\right)1}(t,t^{\prime})$. The sinusoidal
lines represent an excitation, and the dashed lines represent a response.
The blobs in the diagrams represent many-body-particle Green's functions
which need to be simplified by Wick's theorem. As for dotted lines,
they are auxiliary to connect excitations (and responses) with the
same time to the blobs.\label{fig1}}
\end{figure}

\subsection{Linear and nonlinear response}

For a response $R$ excited by an excitation $E$, $V$ , $\rho_{0}^{I}$
and $R$ can be written as the sum of different order of $E$. Here
we take $R(t)$ as example:

\begin{eqnarray}
\hat{R}(t) & = & R_{1}(t)+R_{2}(t)+\cdots\\
 & = & \frac{\partial\hat{R}(t)}{\partial E(t)}E(t)+\frac{1}{2}\frac{\partial^{2}\hat{R}(t)}{\partial E(t)^{2}}E(t)^{2}+\cdots\nonumber \\
 & = & \frac{\partial R_{1}(t)}{\partial E(t)}E(t)+\frac{1}{2}\frac{\partial^{2}R_{2}(t)}{\partial E(t)^{2}}E(t)^{2}+\cdots\nonumber \\
R_{n}(t) & \equiv & \frac{1}{n!}\frac{\partial^{n}R_{n}(t)}{\partial E(t)^{n}}E(t)^{n}.
\end{eqnarray}

And then, the response $\left\langle R(t)\right\rangle =Tr\left[\rho^{I}(t)\hat{R}(t)\right]$
can be written as different orders: 

\begin{figure*}
\includegraphics[width=2\columnwidth]{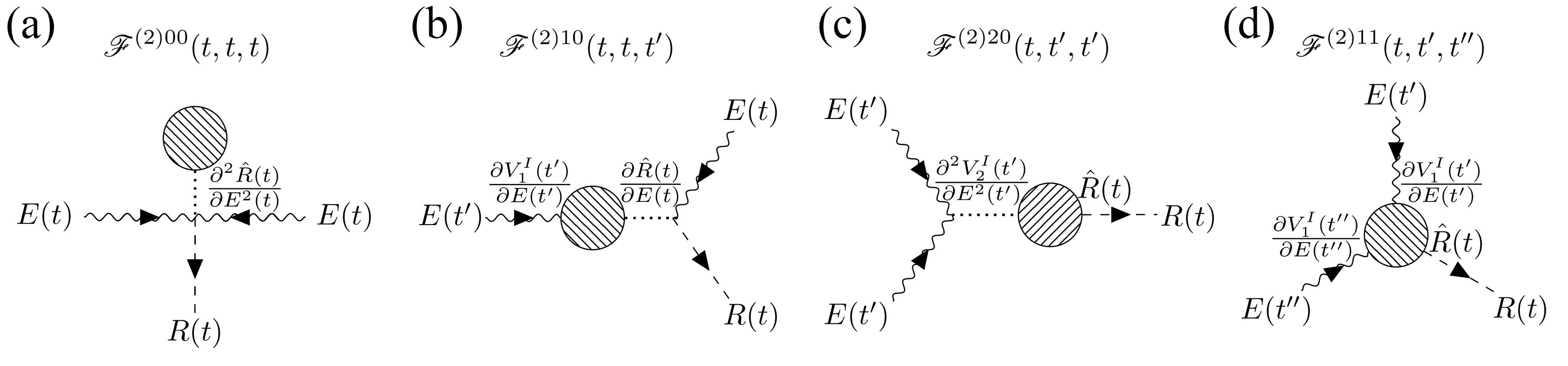} \caption{\textbf{Second-order response Feynman diagrams} (a) $\mathscr{F}^{\left(2\right)00}(t,t,t)$
, (b) $\mathscr{F}^{\left(2\right)10}(t,t,t^{\prime})$, (c) $\mathscr{F}^{\left(2\right)20}(t,t^{\prime},t^{\prime})$
and (d) $\mathscr{F}^{\left(2\right)11}(t,t^{\prime},t^{\prime\prime})$.\label{fig2}}
\end{figure*}

 \begin{widetext}

\begin{eqnarray}
\left\langle R^{(1)}(t)\right\rangle  & = & Tr\left[\rho_{0}^{I}(t)R_{1}(t)\right]+Tr\left[\rho_{1}^{I}(t)R_{0}(t)\right]\\
 & = & Tr\left[\rho_{0}(t)\frac{\partial\hat{R}(t)}{\partial E(t)}\right]E(t)+\int_{-\infty}^{+\infty}dt^{\prime}\Theta(t-t^{\prime})Tr\left[\frac{\partial\rho_{1}^{I}(t)}{\partial V_{1}^{I}(t^{\prime})}\frac{\partial V_{1}^{I}(t^{\prime})}{\partial E(t^{\prime})}\hat{R_{0}}(t)\right]E(t^{\prime})\nonumber 
\end{eqnarray}

\begin{eqnarray}
\left\langle R^{(2)}(t)\right\rangle  & = & Tr\left[\rho_{0}^{I}(t)R_{2}(t)\right]+Tr\left[\rho_{1}^{I}(t)R_{1}(t)\right]+Tr\left[\rho_{2}^{I}(t)R_{0}(t)\right]\\
 & = & +\frac{1}{2}Tr\left[\rho_{0}(t)\frac{\partial^{2}\hat{R}(t)}{\partial E^{2}(t)}\right]E^{2}(t)+\int_{-\infty}^{+\infty}dt^{\prime}\Theta(t-t^{\prime})Tr\left[\frac{\partial\rho_{1}^{I}(t)}{\partial V_{1}^{I}(t^{\prime})}\frac{\partial V_{1}^{I}(t^{\prime})}{\partial E(t^{\prime})}\frac{\partial\hat{R}(t)}{\partial E\left(t\right)}\right]E\left(t\right)E(t^{\prime})\nonumber \\
 &  & +\frac{1}{2}\int_{-\infty}^{+\infty}dt^{\prime}\Theta(t-t^{\prime})Tr\left[\frac{\partial\rho_{2}^{I}(t)}{\partial V_{2}^{I}(t^{\prime})}\frac{\partial^{2}V_{2}^{I}(t^{\prime})}{\partial E^{2}(t^{\prime})}\hat{R}_{0}(t)\right]E^{2}(t^{\prime})\nonumber \\
 &  & +\int\int_{-\infty}^{+\infty}dt^{\prime}dt^{\prime\prime}\Theta(t-t^{\prime})\Theta(t^{\prime}-t^{\prime\prime})Tr\left[\frac{\partial^{2}\rho_{2}^{I}(t)}{\partial V_{1}^{I}(t^{\prime})\partial V_{1}^{I}\left(t^{\prime\prime}\right)}\frac{\partial V_{1}^{I}(t^{\prime})}{\partial E(t^{\prime})}\frac{\partial V_{1}^{I}(t^{\prime\prime})}{\partial E(t^{\prime\prime})}\hat{R}(t)\right]E(t^{\prime})E\left(t^{\prime\prime}\right)\nonumber 
\end{eqnarray}
\begin{eqnarray}
\left\langle R^{(3)}(t)\right\rangle  & = & Tr\left[\rho_{0}^{I}(t)R_{3}(t)\right]+Tr\left[\rho_{1}^{I}(t)R_{2}(t)\right]+Tr\left[\rho_{2}^{I}(t)R_{1}(t)\right]+Tr\left[\rho_{3}^{I}(t)R_{0}(t)\right]\\
 & = & +\frac{1}{3!}Tr\left[\rho_{0}^{I}(t)\frac{\partial^{3}\hat{R}(t)}{\partial E^{3}(t)}\right]E^{3}(t)+\frac{1}{2}\int_{-\infty}^{+\infty}dt^{\prime}\Theta(t-t^{\prime})Tr\left[\frac{\partial\rho_{1}^{I}(t)}{\partial V_{1}^{I}(t^{\prime})}\frac{\partial V_{1}^{I}(t^{\prime})}{\partial E(t^{\prime})}\frac{\partial^{2}\hat{R}(t)}{\partial E^{2}\left(t\right)}\right]E^{2}\left(t\right)E(t^{\prime})\nonumber \\
 &  & +\frac{1}{2}\int_{-\infty}^{+\infty}dt^{\prime}\Theta(t-t^{\prime})Tr\left[\frac{\partial\rho_{2}^{I}(t)}{\partial V_{2}^{I}(t^{\prime})}\frac{\partial^{2}V_{2}^{I}(t^{\prime})}{\partial E^{2}(t^{\prime})}\frac{\partial\hat{R}(t)}{\partial E\left(t\right)}\right]E\left(t\right)E^{2}(t^{\prime})\nonumber \\
 &  & +\int\int_{-\infty}^{+\infty}dt^{\prime}dt^{\prime\prime}\Theta(t-t^{\prime})\Theta(t^{\prime}-t^{\prime\prime})Tr\left[\frac{\partial^{2}\rho_{2}^{I}(t)}{\partial V_{1}^{I}(t^{\prime})\partial V_{1}^{I}\left(t^{\prime\prime}\right)}\frac{\partial V_{1}^{I}(t^{\prime})}{\partial E(t^{\prime})}\frac{\partial V_{1}^{I}(t^{\prime\prime})}{\partial E(t^{\prime\prime})}\frac{\partial\hat{R}(t)}{\partial E\left(t\right)}\right]E(t)E(t^{\prime})E\left(t^{\prime\prime}\right)\nonumber \\
 &  & +\frac{1}{3!}\int_{-\infty}^{+\infty}dt^{\prime}\Theta(t-t^{\prime})Tr\left[\frac{\partial\rho_{3}^{I}(t)}{\partial V_{3}^{I}\left(t^{\prime}\right)}\frac{\partial^{3}V_{3}^{I}\left(t^{\prime}\right)}{\partial E^{3}(t^{\prime})}\hat{R}_{0}(t)\right]E^{3}(t^{\prime})\nonumber \\
 &  & +\frac{1}{2}\int\int_{-\infty}^{+\infty}dt^{\prime}dt^{\prime\prime}\Theta(t-t^{\prime})\Theta(t^{\prime}-t^{\prime\prime})Tr\left[\frac{\partial^{2}\rho_{3}^{I}(t)}{\partial V_{1}^{I}(t^{\prime})\partial V_{2}^{I}\left(t^{\prime\prime}\right)}\frac{\partial V_{1}^{I}(t^{\prime})}{\partial E(t^{\prime})}\frac{\partial^{2}V_{1}^{I}(t^{\prime\prime})}{\partial^{2}E(t^{\prime\prime})}\frac{\partial\hat{R}(t)}{\partial E\left(t\right)}\right]E(t^{\prime})E^{2}\left(t^{\prime\prime}\right)\nonumber \\
 &  & +\frac{1}{2}\int\int_{-\infty}^{+\infty}dt^{\prime}dt^{\prime\prime}\Theta(t-t^{\prime})\Theta(t^{\prime}-t^{\prime\prime})Tr\left[\frac{\partial^{2}\rho_{3}^{I}(t)}{\partial V_{2}^{I}(t^{\prime})\partial V_{1}^{I}\left(t^{\prime\prime}\right)}\frac{\partial^{2}V_{1}^{I}(t^{\prime})}{\partial E^{2}(t^{\prime})}\frac{\partial V_{1}^{I}(t^{\prime\prime})}{\partial E(t^{\prime\prime})}\frac{\partial\hat{R}(t)}{\partial E\left(t\right)}\right]E^{2}(t^{\prime})E\left(t^{\prime\prime}\right)\nonumber \\
 &  & +\int\int\int_{-\infty}^{+\infty}dt^{\prime}dt^{\prime\prime}dt^{\prime\prime\prime}\Theta(t-t^{\prime})\Theta(t^{\prime}-t^{\prime\prime})\Theta(t^{\prime\prime}-t^{\prime\prime\prime})\nonumber \\
 &  & Tr\left[\frac{\partial^{3}\rho_{3}^{I}(t)}{\partial V_{1}^{I}(t^{\prime})\partial V_{1}^{I}\left(t^{\prime\prime}\right)\partial V_{1}^{I}\left(t^{\prime\prime\prime}\right)}\frac{\partial V_{1}^{I}(t^{\prime})}{\partial E(t^{\prime})}\frac{\partial V_{1}^{I}(t^{\prime\prime})}{\partial E(t^{\prime\prime})}\frac{\partial V_{1}^{I}(t^{\prime\prime\prime})}{\partial E(t^{\prime\prime\prime})}\hat{R}(t)\right]E(t^{\prime})E\left(t^{\prime\prime}\right)E(t^{\prime\prime\prime})\nonumber 
\end{eqnarray}

 \end{widetext} 

Further, we can identify response function in different order:

\begin{eqnarray}
\mathscr{F}^{\left(1\right)0}(t,t) & = & Tr\left[\rho_{0}(t)\frac{\partial\hat{R}(t)}{\partial E(t)}\right]\\
 & = & <|\frac{\partial\hat{R}(t)}{\partial E(t)}|>_{0}\nonumber \\
\mathscr{F}^{\left(1\right)1}(t,t^{\prime}) & = & \Theta_{t^{\prime}}^{t}Tr\left[\frac{\partial\rho_{1}^{I}(t)}{\partial V_{1}^{I}(t^{\prime})}\frac{\partial V_{1}^{I}(t^{\prime})}{\partial E(t^{\prime})}\hat{R}(t)\right]\\
 & = & -\frac{i}{\hbar}\Theta_{t^{\prime}}^{t}<|\left[R(t),\frac{\partial V_{1}^{I}(t^{\prime})}{\partial E(t^{\prime})}\right]|>_{0}\nonumber \\
\mathscr{F}^{\left(2\right)00}(t,t,t) & = & \frac{1}{2}Tr\left[\rho_{0}(t)\frac{\partial^{2}\hat{R}(t)}{\partial E^{2}(t)}\right]\\
 & = & \frac{1}{2}<|\frac{\partial^{2}\hat{R}(t)}{\partial E^{2}(t)}|>_{0}\nonumber \\
\mathscr{F}^{\left(2\right)10}(t,t,t^{\prime}) & = & \Theta_{t^{\prime}}^{t}Tr\left[\frac{\partial\rho_{1}^{I}(t)}{\partial V_{1}^{I}(t^{\prime})}\frac{\partial V_{1}^{I}(t^{\prime})}{\partial E(t^{\prime})}\frac{\partial\hat{R}(t)}{\partial E\left(t\right)}\right]\\
 & = & -\frac{i}{\hbar}\Theta_{t^{\prime}}^{t}<|\left[\frac{\partial\hat{R}(t)}{\partial E\left(t\right)},\frac{\partial V_{1}^{I}(t^{\prime})}{\partial E(t^{\prime})}\right]|>_{0}\nonumber \\
\mathscr{F}^{\left(2\right)20}(t,t^{\prime},t^{\prime}) & = & \frac{1}{2}\Theta_{t^{\prime}}^{t}Tr\left[\frac{\partial\rho_{2}(t)}{\partial V_{2}^{I}(t^{\prime})}\frac{\partial^{2}V_{2}^{I}(t^{\prime})}{\partial E^{2}(t^{\prime})}\hat{R}_{0}(t)\right]\\
 & = & -\frac{i}{2\hbar}\Theta_{t^{\prime}}^{t}<|\left[R(t),\frac{\partial^{2}V_{2}^{I}(t^{\prime})}{\partial E^{2}(t^{\prime})}\right]|>_{0}\nonumber 
\end{eqnarray}

\begin{figure*}
\includegraphics[width=2\columnwidth]{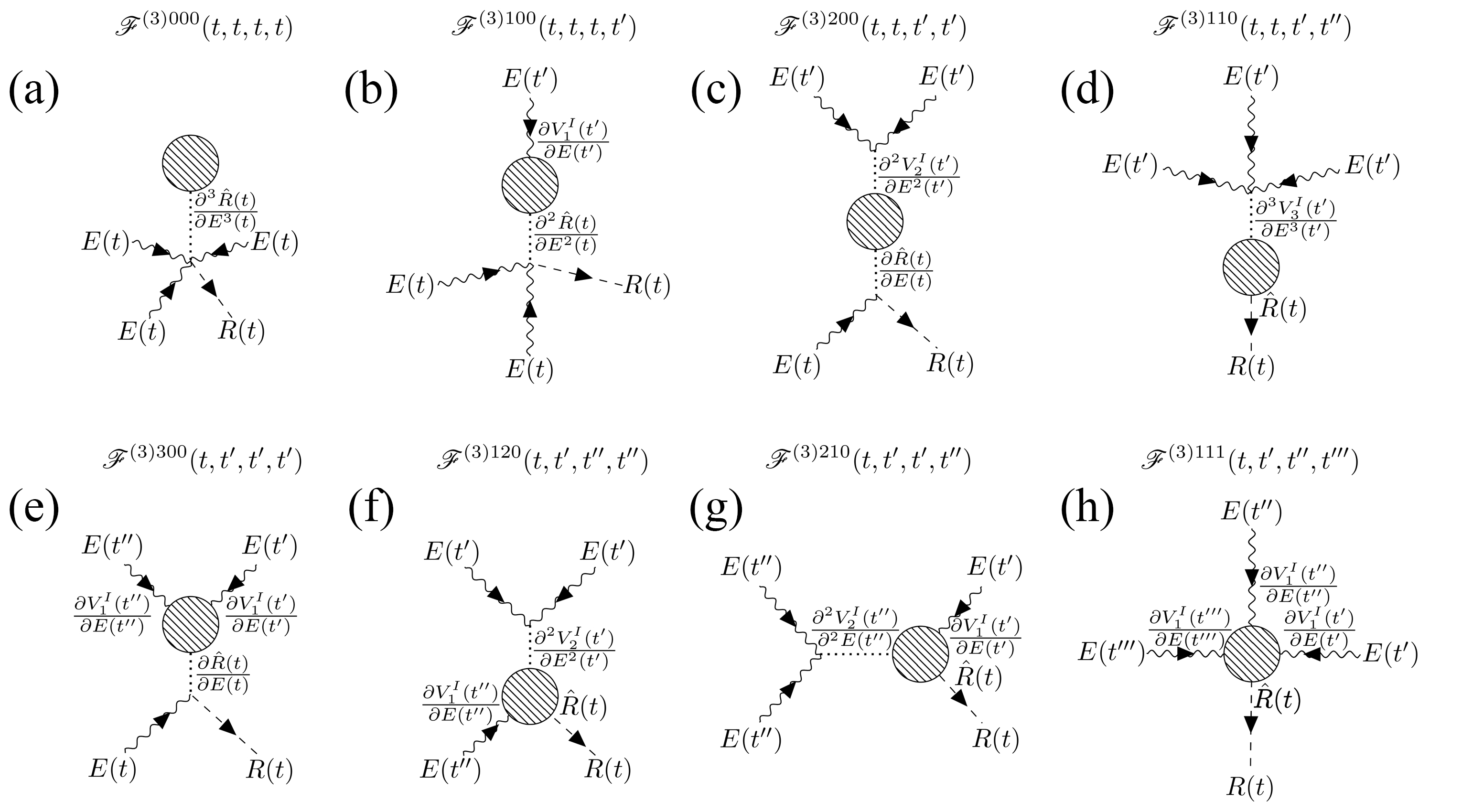} \caption{\textbf{Third-order response Feynman diagrams} (a) $\mathscr{F}^{\left(3\right)000}(t,t,t,t)$,
(b) $\mathscr{F}^{\left(3\right)100}(t,t,t,t^{\prime})$, (c) $\mathscr{F}^{\left(3\right)200}(t,t,t^{\prime},t^{\prime})$,
(d) $\mathscr{F}^{\left(3\right)300}(t,t^{\prime},t^{\prime},t^{\prime})$,
(e) $\mathscr{F}^{\left(3\right)110}(t,t,t^{\prime},t^{\prime\prime})$,
(f) $\mathscr{F}^{\left(3\right)120}(t,t^{\prime},t^{\prime\prime},t^{\prime\prime})$,
(g) $\mathscr{F}^{\left(3\right)210}(t,t^{\prime},t^{\prime},t^{\prime\prime})$
and (h) $\mathscr{F}^{\left(3\right)111}(t,t^{\prime},t^{\prime\prime},t^{\prime\prime\prime})$.\label{fig3}}
\end{figure*}

 \begin{widetext}

\begin{eqnarray}
\mathscr{F}^{\left(2\right)11}(t,t^{\prime},t^{\prime\prime}) & = & \Theta_{t^{\prime}}^{t}\Theta_{t^{\prime\prime}}^{t^{\prime}}Tr\left[\frac{\partial^{2}\rho_{2}^{I}(t)}{\partial V_{1}^{I}(t^{\prime})\partial V_{1}^{I}\left(t^{\prime\prime}\right)}\frac{\partial V_{1}^{I}(t^{\prime})}{\partial E(t^{\prime})}\frac{\partial V_{1}^{I}(t^{\prime\prime})}{\partial E(t^{\prime\prime})}\hat{R}(t)\right]\\
 & = & (-\frac{i}{\hbar})^{2}\Theta_{t^{\prime}}^{t}\Theta_{t^{\prime\prime}}^{t^{\prime}}<|\left[\left[\hat{R}(t),\frac{\partial V_{1}^{I}(t^{\prime})}{\partial E(t^{\prime})}\right],\frac{\partial V_{1}^{I}(t^{\prime\prime})}{\partial E(t^{\prime\prime})}\right]|>_{0}\nonumber \\
\mathscr{F}^{\left(3\right)000}(t,t,t,t) & = & \frac{1}{3!}Tr\left[\rho_{0}(t)\frac{\partial^{3}\hat{R}(t)}{\partial E^{3}(t)}\right]\\
 & = & \frac{1}{3!}<|\frac{\partial^{3}\hat{R}(t)}{\partial E^{3}(t)}|>_{0}\nonumber \\
\mathscr{F}^{\left(3\right)100}(t,t,t,t^{\prime}) & = & \frac{1}{2}\Theta_{t^{\prime}}^{t}Tr\left[\frac{\partial\rho_{1}^{I}(t)}{\partial V_{1}^{I}(t^{\prime})}\frac{\partial V_{1}^{I}(t^{\prime})}{\partial E(t^{\prime})}\frac{\partial^{2}\hat{R}(t)}{\partial E^{2}\left(t\right)}\right]\\
 & = & -\frac{i}{2\hbar}\Theta_{t^{\prime}}^{t}<|\left[\frac{\partial^{2}\hat{R}(t)}{\partial E^{2}\left(t\right)},\frac{\partial V_{1}^{I}(t^{\prime})}{\partial E(t^{\prime})}\right]|>_{0}\nonumber \\
\mathscr{F}^{\left(3\right)200}(t,t,t^{\prime},t^{\prime}) & = & \frac{1}{2}\Theta_{t^{\prime}}^{t}Tr\left[\frac{\partial\rho_{2}^{I}(t)}{\partial V_{2}^{I}(t^{\prime})}\frac{\partial^{2}V_{2}^{I}(t^{\prime})}{\partial E^{2}(t^{\prime})}\frac{\partial\hat{R}(t)}{\partial E\left(t\right)}\right]\\
 & = & -\frac{i}{2\hbar}\Theta_{t^{\prime}}^{t}<|\left[\frac{\partial\hat{R}(t)}{\partial E\left(t\right)},\frac{\partial^{2}V_{2}^{I}(t^{\prime})}{\partial E^{2}(t^{\prime})}\right]|>_{0}\nonumber \\
\mathscr{F}^{\left(3\right)300}(t,t^{\prime},t^{\prime},t^{\prime}) & = & \frac{1}{3!}\Theta_{t^{\prime}}^{t}Tr\left[\frac{\partial\rho_{3}^{I}(t)}{\partial V_{3}^{I}\left(t^{\prime}\right)}\frac{\partial^{3}V_{3}^{I}\left(t^{\prime}\right)}{\partial E^{3}(t^{\prime})}\hat{R}_{0}(t)\right]\\
 & = & -\frac{i}{3!\hbar}\Theta_{t^{\prime}}^{t}<|\left[\hat{R}(t),\frac{\partial^{3}V_{3}^{I}\left(t^{\prime}\right)}{\partial E^{3}(t^{\prime})}\right]|>_{0}\nonumber \\
\mathscr{F}^{\left(3\right)110}(t,t,t^{\prime},t^{\prime\prime}) & = & \Theta_{t^{\prime}}^{t}\Theta_{t^{\prime\prime}}^{t^{\prime}}Tr\left[\frac{\partial^{2}\rho_{2}^{I}(t)}{\partial V_{1}^{I}(t^{\prime})\partial V_{1}^{I}\left(t^{\prime\prime}\right)}\frac{\partial V_{1}^{I}(t^{\prime})}{\partial E(t^{\prime})}\frac{\partial V_{1}^{I}(t^{\prime\prime})}{\partial E(t^{\prime\prime})}\frac{\partial\hat{R}(t)}{\partial E\left(t\right)}\right]\\
 & = & (-\frac{i}{\hbar})^{2}\Theta_{t^{\prime}}^{t}\Theta_{t^{\prime\prime}}^{t^{\prime}}<|\left[\left[\frac{\partial\hat{R}(t)}{\partial E\left(t\right)},\frac{\partial V_{1}^{I}(t^{\prime})}{\partial E(t^{\prime})}\right],\frac{\partial V_{1}^{I}(t^{\prime\prime})}{\partial E(t^{\prime\prime})}\right]|>_{0}\nonumber \\
\mathscr{F}^{\left(3\right)120}(t,t^{\prime},t^{\prime\prime},t^{\prime\prime}) & = & \frac{1}{2}\Theta_{t^{\prime}}^{t}\Theta_{t^{\prime\prime}}^{t^{\prime}}Tr\left[\frac{\partial^{2}\rho_{3}^{I}(t)}{\partial V_{1}^{I}(t^{\prime})\partial V_{2}^{I}\left(t^{\prime\prime}\right)}\frac{\partial V_{1}^{I}(t^{\prime})}{\partial E(t^{\prime})}\frac{\partial^{2}V_{1}^{I}(t^{\prime\prime})}{\partial^{2}E(t^{\prime\prime})}\frac{\partial\hat{R}(t)}{\partial E\left(t\right)}\right]\\
 & = & \frac{1}{2}(-\frac{i}{\hbar})^{2}\Theta_{t^{\prime}}^{t}\Theta_{t^{\prime\prime}}^{t^{\prime}}<|\left[\left[\hat{R}(t),\frac{\partial V_{1}^{I}(t^{\prime})}{\partial E(t^{\prime})}\right],\frac{\partial^{2}V_{1}^{I}(t^{\prime\prime})}{\partial^{2}E(t^{\prime\prime})}\right]|>_{0}\nonumber \\
\mathscr{F}^{\left(3\right)210}(t,t^{\prime},t^{\prime},t^{\prime\prime}) & = & \frac{1}{2}\Theta_{t^{\prime}}^{t}\Theta_{t^{\prime\prime}}^{t^{\prime}}Tr\left[\frac{\partial^{2}\rho_{3}^{I}(t)}{\partial V_{2}^{I}(t^{\prime})\partial V_{1}^{I}\left(t^{\prime\prime}\right)}\frac{\partial^{2}V_{1}^{I}(t^{\prime})}{\partial E^{2}(t^{\prime})}\frac{\partial V_{1}^{I}(t^{\prime\prime})}{\partial E(t^{\prime\prime})}\frac{\partial\hat{R}(t)}{\partial E\left(t\right)}\right]\\
 & = & \frac{1}{2}(-\frac{i}{\hbar})^{2}\Theta_{t^{\prime}}^{t}\Theta_{t^{\prime\prime}}^{t^{\prime}}<|\left[\left[\hat{R}(t),\frac{\partial^{2}V_{1}^{I}(t^{\prime})}{\partial E^{2}(t^{\prime})}\right],\frac{\partial V_{1}^{I}(t^{\prime\prime})}{\partial E(t^{\prime\prime})}\right]|>_{0}\nonumber \\
\mathscr{F}^{\left(3\right)111}(t,t^{\prime},t^{\prime\prime},t^{\prime\prime\prime}) & = & \Theta_{t^{\prime}}^{t}\Theta_{t^{\prime\prime}}^{t^{\prime}}\Theta_{t^{\prime\prime\prime}}^{t^{\prime\prime}}Tr\left[\frac{\partial^{3}\rho_{3}^{I}(t)}{\partial V_{1}^{I}(t^{\prime})\partial V_{1}^{I}\left(t^{\prime\prime}\right)\partial V_{1}^{I}\left(t^{\prime\prime\prime}\right)}\frac{\partial V_{1}^{I}(t^{\prime})}{\partial E(t^{\prime})}\frac{\partial V_{1}^{I}(t^{\prime\prime})}{\partial E(t^{\prime\prime})}\frac{\partial V_{1}^{I}(t^{\prime\prime\prime})}{\partial E(t^{\prime\prime\prime})}\hat{R}(t)\right]\\
 & = & (-\frac{i}{\hbar})^{3}\Theta_{t^{\prime}}^{t}\Theta_{t^{\prime\prime}}^{t^{\prime}}\Theta_{t^{\prime\prime\prime}}^{t^{\prime\prime}}<|\left[\left[\left[\hat{R}(t),\frac{\partial V_{1}^{I}(t^{\prime})}{\partial E(t^{\prime})}\right],\frac{\partial V_{1}^{I}(t^{\prime\prime})}{\partial E(t^{\prime\prime})}\right],\frac{\partial V_{1}^{I}(t^{\prime\prime\prime})}{\partial E(t^{\prime\prime\prime})}\right]|>_{0}\nonumber 
\end{eqnarray}
 \end{widetext}The first number in the bracket of response function
is marked the order of the response and the rest of numbers are marked
the number of $t^{\prime}$, $t^{\prime\prime}$ and so on. As for
step function $\Theta(t-t^{\prime})$, we use $\Theta_{t^{\prime}}^{t}$for
simplification. Here we can see that for n-th order responses, there
are $2^{n}$ terms response functions including all situations. We
can also consider in-homogeneous perturbation $V(r,t)$ to study more
complicated system.

\section{Optical current in different orders}

For example, if we consider optical current with the Hamiltonian

\begin{eqnarray}
H_{k}^{A} & = & \frac{\left(\boldsymbol{p}+\hbar k+\boldsymbol{eA}\right)^{2}}{2m}+V^{L}\left(r\right)+V^{NL}(\boldsymbol{p}+\hbar k+\boldsymbol{eA})\nonumber \\
\end{eqnarray}
($H_{k}^{A}$ can be got from $H_{k}$ using the replacement $\hbar k\rightarrow\hbar k+\boldsymbol{eA}$),
the perturbation $V$ inside could be written in different order 
\begin{eqnarray}
V & = & \sum_{n=1}^{\infty}\frac{\boldsymbol{e}^{n}\partial^{n}H_{k}}{n!\hbar^{n}\partial k^{b}\partial k^{c}\cdots}A_{b}A_{c}\cdots\\
 & = & \sum_{n=1}^{\infty}\frac{e^{-i(\omega_{1}+\omega_{2}+\cdots)t}\boldsymbol{e}^{n}\partial^{n}H_{k}}{n!(i\hbar)^{n}\omega_{1}\partial k^{b}\omega_{2}\partial k^{c}\cdots}\mathbf{E}_{b}\left(\omega_{1}\right)\mathbf{E}_{c}\left(\omega_{2}\right)\cdots\nonumber \\
 & = & \sum_{n=1}^{\infty}\frac{e^{-i(\omega_{1}+\omega_{2}+\cdots)t}\boldsymbol{e}^{n}}{i^{n}n!\omega_{1}\omega_{2}\cdots}v^{bc\cdots}\mathbf{E}_{b}\left(\omega_{1}\right)\mathbf{E}_{c}\left(\omega_{2}\right)\cdots\nonumber 
\end{eqnarray}
 ($v^{bc\cdots}=\frac{\partial^{n}H_{k}}{\hbar^{n}\partial k^{b}\partial k^{c}\cdots}$),
the excitation is electric field $\mathbf{E}\left(\omega\right)$,($\boldsymbol{A}=\frac{\tilde{\mathbf{E}}_{\omega}}{i\omega}$,
$\tilde{\mathbf{E}}_{\omega}\left(t\right)=\mathbf{E}\left(\omega\right)e^{-i\omega t}$)
, and the response is 
\begin{eqnarray}
J_{a}\left(\omega\right) & = & \frac{-\boldsymbol{e}\partial H_{k}^{A}}{\hbar\partial k^{a}}\\
 & = & -\boldsymbol{e}v^{a}+\nonumber \\
 &  & \sum_{n=1}^{\infty}\frac{e^{-i(\omega_{1}+\omega_{2}+\cdots)t}\left(-\boldsymbol{e}^{n+1}\right)v^{abc\cdots}\mathbf{E}_{b}\left(\omega_{b}\right)\mathbf{E}_{c}\left(\omega_{c}\right)\cdots}{i^{n}n!\omega_{1}\omega_{2}\cdots}\nonumber 
\end{eqnarray}

Take the perturbation $V$, the excitation $\mathbf{E}\left(\omega\right)$
and the response $J_{a}\left(\omega\right)$ inside the response function
before, we can get conductivities in different orders.

\subsection{Linear optical conductivity, Berry curvature}

Here is linear conductivities as function of time:
\begin{eqnarray}
\mathscr{\sigma}_{ab}^{\left(1\right)0}(t,t) & = & \int d\widetilde{\boldsymbol{k}}<|\frac{\partial\hat{R}^{a}(t)}{\partial\mathbf{E}^{b}(t)}|>_{0}\\
 & = & \frac{\left(-\boldsymbol{e}^{2}\right)e^{-i\omega_{1}t}}{i\omega_{1}}\int d\widetilde{\boldsymbol{k}}<|v^{ab}|>_{0}\nonumber \\
\mathscr{\sigma}_{ab}^{\left(1\right)1}(t,t^{\prime}) & = & -\frac{i}{\hbar}\int d\widetilde{\boldsymbol{k}}\Theta_{t^{\prime}}^{t}<|\left[R(t),\frac{\partial V_{1}^{I}(t^{\prime})}{\partial E(t^{\prime})}\right]|>_{0}\\
 & = & -\frac{\left(-\boldsymbol{e}^{2}\right)}{\hbar\omega_{1}}\int d\widetilde{\boldsymbol{k}}\Theta_{t^{\prime}}^{t}e^{-i\omega_{1}t^{\prime}}<|\left[v^{a}\left(t\right),v^{b}\left(t^{\prime}\right)\right]|>_{0}\nonumber 
\end{eqnarray}

Generally, we do a Fourier transform for conductivities to switch
it to frequency-dependent one

\begin{eqnarray}
 &  & \mathscr{\sigma}_{ab}^{\left(1\right)0}(\omega=\omega_{1})\\
 & = & \frac{1}{2\pi}\int_{-\infty}^{+\infty}dte^{i\omega t}\mathscr{\sigma}_{ab}^{\left(1\right)0}(t,t)\nonumber \\
 & = & \frac{\left(-\boldsymbol{e}^{2}\right)}{i\omega}\int d\widetilde{\boldsymbol{k}}<|v^{ab}|>_{0}\nonumber \\
 & = & \left(-\frac{\boldsymbol{e}^{2}}{\hbar}\right)\frac{\hbar}{\omega}\iint d\widetilde{\boldsymbol{k}}d\widetilde{E}f(E)Tr\left(v^{ab}g_{E}^{r-a}\right)\nonumber \\
 & = & \left(-\frac{\boldsymbol{e}^{2}}{\hbar}\right)\frac{\hbar}{i\omega}\int d\widetilde{\boldsymbol{k}}\sum_{n}f(\varepsilon_{nk})v_{nn}^{ab}\nonumber 
\end{eqnarray}

\begin{eqnarray}
 &  & \mathscr{\sigma}_{ab}^{\left(1\right)1}(\omega=\omega_{1})\\
 & = & \frac{1}{2\pi}\iint_{-\infty}^{+\infty}dtdt^{\prime}e^{i\omega t}\mathscr{\sigma}_{ab}^{\left(1\right)1}(t,t^{\prime})\nonumber \\
 & = & \left(-\frac{\boldsymbol{e}^{2}}{\hbar}\right)\frac{\hbar}{\omega}\iint d\widetilde{\boldsymbol{k}}d\widetilde{E}f(E)\times\nonumber \\
 &  & \,\,\,\,Tr\left(v^{a}g_{E+\hbar\omega}^{r}v^{b}g_{E}^{r-a}+g_{E}^{r-a}v^{b}g_{E-\hbar\omega}^{a}v^{a}\right)\nonumber \\
 & = & \left(-\frac{\boldsymbol{e}^{2}}{\hbar}\right)\frac{\hbar}{i\omega}\int d\widetilde{\boldsymbol{k}}\sum_{nm}f_{nm}\frac{v_{nm}^{a}v_{mn}^{b}}{\varepsilon_{nk}-\varepsilon_{mk}+\hbar\omega+i\eta},\nonumber 
\end{eqnarray}
where $\int d\widetilde{\boldsymbol{k}}=\frac{1}{(2\pi)^{d}}\int d\boldsymbol{k}=\underset{N_{k}\rightarrow\infty}{lim}\frac{1}{L^{d}N_{k}}\underset{\boldsymbol{k}}{\sum}$,
$\int d\widetilde{E}=\int\frac{dE}{2\pi}$, $g_{E}^{r}$ is retarded
green's function $g^{r}(E)$ , $g_{E}^{r-a}=g^{r}(E)-g^{a}(E)$ and
$f_{nm}=f(\varepsilon_{nk})-f(\varepsilon_{mk})$ (seeing Appendix
\ref{sec:Aa}\&\ref{sec:Ab} for details).

In the one hand, for first order response here, $\mathscr{\mathscr{\sigma}}_{ab}^{10}(\omega)$
is Drude term. If we consider a free-electron-gas system with the
Hamiltonian $H=\frac{\hbar^{2}k^{2}}{2m^{\star}}$ at zero frequency
limit $\left(\left.\omega\rightarrow\left(\omega+\frac{i}{\tau}\right)\right|_{\omega\rightarrow0}\right)$
, we can get classical longitudinal conductivity $\sigma_{xx}=\frac{ne^{2}\tau}{m^{\star}}$.
On the other hand, for transverse conductivity at zero frequency limit,
it describes anomalous Hall effect:

\begin{eqnarray}
\sigma_{ab} & = & \frac{1}{2}\left[\mathscr{\sigma}_{ab}^{\left(1\right)1}(\omega)+\mathscr{\sigma}_{ab}^{\left(1\right)1}(-\omega)\right]|_{\omega\rightarrow0}\label{eq:berrycur}\\
 & = & \left(-\frac{\boldsymbol{e}^{2}}{\hbar}\right)\frac{\hbar^{2}}{i}\int d\widetilde{\boldsymbol{k}}\sum_{nm}f_{nm}\frac{v_{nm}^{a}v_{mn}^{b}}{\left(\varepsilon_{nk}-\varepsilon_{mk}\right)^{2}}\nonumber \\
 & = & \left(-\frac{\boldsymbol{e}^{2}}{\hbar}\right)\hbar^{2}\iint d\widetilde{\boldsymbol{k}}d\widetilde{E}f(E)\times\nonumber \\
 &  & \,\,\,\,\,\,\,\,\,\,\,Tr\left(v^{a}\frac{\partial g_{E}^{r}}{\partial E}v^{b}g_{E}^{r-a}-g_{E}^{r-a}v^{b}\frac{\partial g_{E}^{a}}{\partial E}v^{a}\right),\nonumber 
\end{eqnarray}
which is the integral of Berry curvature. In previous works of Ref.\citep{potp2006-nagaosa-nonequilibirum,prb2008-nagaosa-nonequilibirum,RMP-AHE-nagaosa},
the Green's function formulas of anomalous Hall effect resulted from
Berry curvature is got based on the Keldysh formalism in the gauge-covariant
Wigner space. For terms in Eq.\ref{eq:berrycur}, we only have Fermi-sea
contribution while there are both Fermi-surface and Fermi-sea terms
in Ref.\citep{potp2006-nagaosa-nonequilibirum,prb2008-nagaosa-nonequilibirum,RMP-AHE-nagaosa}.
Using integration by parts of energy $E$, we can also have several
terms with the contribution of Fermi-surface $\int dE\frac{\partial f(E)}{\partial E}=-\int dE\delta(E-\mu)$
and Fermi-sea $\int dEf(E)$ terms at limit of zero temperature.

\subsection{Second-order nonlinear optical conductivity, shift current, Berry
curvature dipole}

Based on similar method, we can deal with the second order response:
\begin{eqnarray}
 &  & \mathscr{\sigma}_{abc}^{\left(2\right)00}(t,t,t)\\
 & = & -\frac{\left(-\boldsymbol{e}^{3}\right)e^{-i\omega_{1}t-i\omega_{2}t}}{2\omega_{1}\omega_{2}}\int d\widetilde{\boldsymbol{k}}<|v^{abc}|>_{0}\nonumber 
\end{eqnarray}
\begin{eqnarray}
 &  & \mathscr{\sigma}_{abc}^{200}(\omega=\omega_{1}+\omega_{2})\\
 & = & -\frac{\left(-\boldsymbol{e}^{3}\right)}{2\omega_{1}\omega_{2}}<|v^{abc}|>_{0}\nonumber \\
 & = & -\left(-\frac{\boldsymbol{e}^{3}}{\hbar^{2}}\right)\frac{i\hbar^{2}}{\omega_{1}\omega_{2}}\iint d\widetilde{\boldsymbol{k}}d\widetilde{E}f(E)Tr\left(v^{abc}g_{E}^{r-a}\right)\nonumber \\
 & = & -\left(-\frac{\boldsymbol{e}^{3}}{\hbar^{2}}\right)\frac{\hbar^{2}}{\omega_{1}\omega_{2}}\int d\widetilde{\boldsymbol{k}}\sum_{n}f(\varepsilon_{nk})v_{nn}^{abc}\nonumber 
\end{eqnarray}

\begin{eqnarray}
 &  & \mathscr{\sigma}_{abc}^{\left(2\right)10}(t,t,t^{\prime})\\
 & = & \frac{i\left(-\boldsymbol{e}^{3}\right)e^{-i\omega_{1}t-i\omega_{2}t^{\prime}}}{\hbar\omega_{1}\omega_{2}}\Theta_{t^{\prime}}^{t}\int d\widetilde{\boldsymbol{k}}<|\left[v^{ab}(t),v^{c}\left(t^{\prime}\right)\right]|>_{0}\nonumber 
\end{eqnarray}

\begin{eqnarray}
 &  & \mathscr{\sigma}_{abc}^{\left(2\right)10}(\omega=\omega_{1}+\omega_{2})\\
 & = & -\left(-\frac{\boldsymbol{e}^{3}}{\hbar^{2}}\right)\frac{i\hbar^{2}}{\omega_{1}\omega_{2}}\iint d\widetilde{\boldsymbol{k}}d\widetilde{E}f(E)\times\nonumber \\
 &  & \,\,\,\,\,\,\,\,\,Tr\left(v^{ab}g_{E+\hbar\omega_{2}}^{r}v^{c}g_{E}^{r-a}+g_{E}^{r-a}v^{c}g_{E-\hbar\omega_{2}}^{a}v^{ab}\right)\nonumber \\
 & = & -\left(-\frac{\boldsymbol{e}^{3}}{\hbar^{2}}\right)\frac{\hbar^{2}}{\omega_{1}\omega_{2}}\int d\widetilde{\boldsymbol{k}}\sum_{nm}f_{nm}\frac{v_{nm}^{ab}v_{mn}^{c}}{\varepsilon_{nk}-\varepsilon_{mk}+\hbar\omega_{2}+i\eta}\nonumber 
\end{eqnarray}

\begin{eqnarray}
 &  & \mathscr{\sigma}_{abc}^{\left(2\right)20}(t,t^{\prime},t^{\prime})\\
 & = & \frac{i\left(-\boldsymbol{e}^{3}\right)e^{-i\omega_{1}t^{\prime}}e^{-i\omega_{2}t^{\prime}}}{2\hbar\omega_{1}\omega_{2}}\Theta_{t^{\prime}}^{t}\int d\widetilde{\boldsymbol{k}}<|\left[v^{a}\left(t\right),v^{bc}\left(t^{\prime}\right)\right]|>_{0}\nonumber 
\end{eqnarray}

\begin{eqnarray}
 &  & \mathscr{\sigma}_{abc}^{\left(2\right)20}(\omega=\omega_{1}+\omega_{2})\\
 & = & -\left(-\frac{\boldsymbol{e}^{3}}{\hbar^{2}}\right)\frac{i\hbar^{2}}{2\omega_{1}\omega_{2}}\iint d\widetilde{\boldsymbol{k}}d\widetilde{E}f(E)\times\nonumber \\
 &  & \,\,\,\,\,\,\,\,Tr\left(v^{a}g_{E+\hbar\omega}^{r}v^{bc}g_{E}^{r-a}+g_{E}^{r-a}v^{bc}g_{E-\hbar\omega}^{a}v^{a}\right)\nonumber \\
 & = & -\left(-\frac{\boldsymbol{e}^{3}}{\hbar^{2}}\right)\frac{\hbar^{2}}{2\omega_{1}\omega_{2}}\int d\widetilde{\boldsymbol{k}}\sum_{nm}f_{nm}\frac{v_{nm}^{a}v_{mn}^{bc}}{\varepsilon_{nk}-\varepsilon_{mk}+\hbar\omega+i\eta}\nonumber 
\end{eqnarray}

\begin{eqnarray}
\mathscr{\sigma}_{abc}^{\left(2\right)11}(t,t^{\prime},t^{\prime\prime}) & = & \frac{\left(-\boldsymbol{e}^{3}\right)e^{-i\omega_{1}t^{\prime}-i\omega_{2}t^{\prime\prime}}}{\hbar^{2}\omega_{1}\omega_{2}}\Theta_{t^{\prime}}^{t}\Theta_{t^{\prime\prime}}^{t^{\prime}}\times\\
 &  & \int d\widetilde{\boldsymbol{k}}<|\left[\left[v^{a}\left(t\right),v^{b}\left(t^{\prime}\right)\right],v^{c}\left(t^{\prime\prime}\right)\right]|>_{0}\nonumber 
\end{eqnarray}

\begin{eqnarray}
 &  & \mathscr{\sigma}_{abc}^{\left(2\right)11}(\omega=\omega_{1}+\omega_{2})\\
 & = & \,\,\,\left(-\frac{\boldsymbol{e}^{3}}{\hbar^{2}}\right)\frac{\hbar^{2}}{\omega_{1}\omega_{2}}\iiint d\widetilde{\boldsymbol{k}}d\widetilde{E}d\widetilde{E}^{\prime}\left[f(E)-f(E^{\prime})\right]\times\nonumber \\
 &  & \,\,\,\,\,\,\,\,\,\,\,\,\,\,\,\,\,\,\,\,\,\,\,\,\,\,\,Tr\left(\frac{v^{a}g_{E+\hbar\omega}^{r}v^{b}g_{E^{\prime}}^{r-a}v^{c}g_{E}^{r-a}}{E+\hbar\omega_{2}-E^{\prime}+i\eta}\right.\nonumber \\
 &  & \,\,\,\,\,\,\,\,\,\,\,\,\,\,\,\,\,\,\,\,\,\,\,\,\,\,\,\,\,\,\left.+\frac{g_{E}^{r-a}v^{c}g_{E^{\prime}}^{r-a}v^{b}g_{E-\hbar\omega}^{a}v^{a}}{E-\hbar\omega_{2}-E^{\prime}-i\eta}\right)\nonumber \\
 & = & \left(-\frac{\boldsymbol{e}^{3}}{\hbar^{2}}\right)\frac{\hbar^{2}}{\omega_{1}\omega_{2}}\int d\widetilde{\boldsymbol{k}}\sum_{lmn}f_{nl}\times\nonumber \\
 &  & \,\,\,\,\left[\frac{v_{lm}^{a}v_{mn}^{b}v_{nl}^{c}}{\left(\varepsilon_{lk}+\hbar\omega_{2}-\varepsilon_{nk}+i\eta\right)\left(\varepsilon_{lk}+\hbar\omega-\varepsilon_{mk}+i\eta\right)}\right.\nonumber \\
 &  & +\left.\frac{v_{ln}^{c}v_{nm}^{b}v_{ml}^{a}}{\left(\varepsilon_{lk}-\hbar\omega_{2}-\varepsilon_{nk}-i\eta\right)\left(\varepsilon_{lk}-\hbar\omega-\varepsilon_{mk}-i\eta\right)}\right]\nonumber 
\end{eqnarray}
(seeing Appendix \ref{sec:Aa}\&\ref{sec:Ab} for details).

In second-order response, it shows more novel physical properties
since it comes from double perturbation. In optical current researches,
two photons of $\omega_{1}$, $\omega_{2}$with different polarization
cases has abundant effect. Recently, for example, shift current has
been widely discussed in several materials. It usually describes Bloch
electron shift some distance after transitions between valence band
and conduction band perturbed by two photons with the same linear
polarization, resulting in direct current in the system. Because it
involves interband process, optical conductivity of shift current
$\bar{\sigma}_{abb,shift}^{\left(2\right)}$ should not include $\mathscr{\sigma}_{abb}^{\left(2\right)00}$:

\begin{eqnarray}
\bar{\sigma}_{abb,shift}^{\left(2\right)} & = & \frac{1}{2}\underset{\alpha\neq00}{\sum}\left[\mathscr{\sigma}_{abb}^{\left(2\right)\alpha}(\omega,-\omega)+\mathscr{\sigma}_{abb}^{\left(2\right)\alpha}(-\omega,\omega)\right]\nonumber \\
 & = & \underset{\alpha\neq00}{\sum}\mathrm{Re}\left[\mathscr{\sigma}_{abb}^{\left(2\right)\alpha}(\omega,-\omega)\right]\,,
\end{eqnarray}
where in our framework of conductivies, it usually has the relationship
of $\mathscr{\sigma}_{abb}^{\left(2\right)\alpha}(\omega,-\omega)=\left[\mathscr{\sigma}_{abb}^{\left(2\right)\alpha}(-\omega,\omega)\right]^{*}$.

As for the case of circular photogalvanic effect ( CPGE ), it describes
optical direct current generated by circular polarized photons. As
we know, one circular polarized light can be written as the sum of
two linear polarized lights with orthogonal polarization, so its optical
conductivity could be written as $\bar{\sigma}_{abc,CPGE}^{\left(2\right)}$,
where $b\neq c$. 

\begin{eqnarray}
 &  & \bar{\sigma}_{abc,CPGE}^{\left(2\right)}\label{eq:intraBCD-1}\\
 & = & \frac{1}{2}\underset{\alpha=10,11}{\sum}\left[\mathscr{\sigma}_{abc}^{\left(2\right)\alpha}(\omega,-\omega)-\mathscr{\sigma}_{abc}^{\left(2\right)\alpha}(-\omega,\omega)\right.\nonumber \\
 &  & \left.-\mathscr{\sigma}_{acb}^{\left(2\right)\alpha}(\omega,-\omega)+\mathscr{\sigma}_{acb}^{\left(2\right)\alpha}(-\omega,\omega)\right]\nonumber \\
 & = & \left(-\frac{\boldsymbol{e}^{3}}{\hbar^{2}}\right)\frac{\hbar^{2}}{\omega}\int d\widetilde{\boldsymbol{k}}\sum_{nm}f_{nm}\frac{v_{nm}^{ab}v_{mn}^{c}-v_{nm}^{ac}v_{mn}^{b}}{\left(\varepsilon_{nk}-\varepsilon_{mk}+i\eta\right)^{2}-\left(\hbar\omega\right)^{2}}\nonumber \\
 &  & -\left(-\frac{\boldsymbol{e}^{3}}{\hbar^{2}}\right)\frac{\hbar^{2}}{\omega}\int d\widetilde{\boldsymbol{k}}\sum_{lmn}f_{nl}\times\nonumber \\
 &  & \left[\frac{v_{lm}^{a}v_{mn}^{b}v_{nl}^{c}-v_{lm}^{a}v_{mn}^{c}v_{nl}^{b}}{\left[\left(\varepsilon_{lk}-\varepsilon_{nk}+i\eta\right)^{2}-\left(\hbar\omega\right)^{2}\right]\left(\varepsilon_{lk}-\varepsilon_{mk}+i\eta\right)}\right.\nonumber \\
 &  & -\left.\frac{v_{ln}^{c}v_{nm}^{b}v_{ml}^{a}-v_{ln}^{b}v_{nm}^{c}v_{ml}^{a}}{\left[\left(\varepsilon_{lk}-\varepsilon_{nk}-i\eta\right)^{2}-\left(\hbar\omega\right)^{2}\right]\left(\varepsilon_{lk}-\varepsilon_{mk}-i\eta\right)}\right]\nonumber \\
 & = & -\left(-\frac{\boldsymbol{e}^{3}}{\hbar^{2}}\right)\frac{i\hbar^{2}}{\omega}\mathrm{Re}\iint d\widetilde{\boldsymbol{k}}d\widetilde{E}f(E)\times\nonumber \\
 &  & \,\,\,\,\,\,\,\,\,Tr\left(v^{ab}\frac{\partial g_{E}^{r}}{\partial E}v^{c}g_{E}^{r-a}-v^{ac}\frac{\partial g_{E}^{r}}{\partial E}v^{b}g_{E}^{r-a}\right)\nonumber \\
 &  & \,\,\,\left(-\frac{\boldsymbol{e}^{3}}{\hbar^{2}}\right)\frac{\hbar^{2}}{\omega}\mathrm{Re}\iiint d\widetilde{\boldsymbol{k}}d\widetilde{E}d\widetilde{E}^{\prime}\left[f(E)-f(E^{\prime})\right]\times\nonumber \\
 &  & \,\,\,\,\,Tr\left[\frac{v^{a}g_{E}^{r}v^{b}g_{E^{\prime}}^{r-a}v^{c}g_{E}^{r-a}-v^{a}g_{E}^{r}v^{c}g_{E^{\prime}}^{r-a}v^{b}g_{E}^{r-a}}{\left(E-E^{\prime}+i\eta\right)^{2}-\left(\hbar\omega\right)^{2}}\right]\nonumber 
\end{eqnarray}

According to previous Ref, at limit of low frequent limit $\omega\rightarrow0$
and $\omega\tau\gg1$, $\bar{\sigma}_{abc,CPGE}^{\left(2\right)}\varpropto\frac{D^{abc}-D^{acb}}{i\omega}$,
where $D^{abc}$ is called Berry curvature dipole, or intraband Berry
curvature dipole. So for Eq.\ref{eq:intraBCD-1}, it could be also
regarded as the Green's function formulas for intraband Berry curvature
dipole (seeing more details in Appendix \ref{sec:Ac}).

We can also derive the form of Green's function formulas for interband
Berry curvature dipole:

\begin{eqnarray}
 &  & \bar{\sigma}_{D,inter}^{\left(2\right)}\\
 & = & \frac{1}{2}\underset{\alpha=10,11}{\sum}\left[\mathscr{\sigma}_{abc}^{\left(2\right)\alpha}(\omega,-\omega)+\mathscr{\sigma}_{abc}^{\left(2\right)\alpha}(-\omega,\omega)\right.\nonumber \\
 &  & \left.+\mathscr{\sigma}_{acb}^{\left(2\right)\alpha}(\omega,-\omega)+\mathscr{\sigma}_{acb}^{\left(2\right)\alpha}(-\omega,\omega)\right]\nonumber \\
 & = & \left(-\frac{\boldsymbol{e}^{3}}{\hbar^{2}}\right)\frac{\hbar^{2}}{2\omega^{2}}\mathrm{Im}\iint d\widetilde{\boldsymbol{k}}d\widetilde{E}f(E)\times\nonumber \\
 &  & \,\,\,\,\,\,\,\,\,Tr\left[v^{ab}\left(g_{E+\hbar\omega}^{r}+g_{E-\hbar\omega}^{r}\right)v^{c}g_{E}^{r-a}\right.\nonumber \\
 &  & \,\,\,\,\,\,\,\,\,\,\left.+v^{ac}\left(g_{E+\hbar\omega}^{r}+g_{E-\hbar\omega}^{r}\right)v^{b}g_{E}^{r-a}\right]\nonumber \\
 &  & -\left(-\frac{\boldsymbol{e}^{3}}{\hbar^{2}}\right)\frac{\hbar^{2}}{\omega^{2}}\mathrm{Re}\iiint d\widetilde{\boldsymbol{k}}d\widetilde{E}d\widetilde{E}^{\prime}\times\nonumber \\
 &  & \,\,\,\,\,\,\,\,\,\,\,\,\,\,\,\,\,\,\,\,\,\,\,\,\frac{\left[f(E)-f(E^{\prime})\right]\left(E-E^{\prime}+i\eta\right)}{\left(E-E^{\prime}+i\eta\right)^{2}-\left(\hbar\omega\right)^{2}}\times\nonumber \\
 &  & \,\,\,\,\,\,\,\,\,\,\,\,\,\,\,\,\,\,\,\,\,Tr\left[v^{a}g_{E}^{r}\left(v^{b}g_{E^{\prime}}^{r-a}v^{c}+v^{c}g_{E^{\prime}}^{r-a}v^{b}\right)g_{E}^{r-a}\right]\,.\nonumber 
\end{eqnarray}
Here, $b$ and $c$ also represent orthogonal polarization, and the
electric fields of direction b and c have the same phases, instead
of $\frac{\pi}{2}$ phase difference in CPGE. It includes the term
of $\delta(E-E^{\prime}\pm\hbar\omega)$ in $\frac{\left(E-E^{\prime}+i\eta\right)}{\left(E-E^{\prime}+i\eta\right)^{2}-\left(\hbar\omega\right)^{2}}$
or spectral function $\mathrm{Im}\left(g_{E\pm\hbar\omega}^{r}\right)$,
which is the origin of the name $interband$ Berry curvature dipole.

\subsection{Third-order optical response, Nonlinear Hall Effect}

Nonlinear optical conductivity in third order also attracts attention.
For third order responses in optical conductivity, we have eight terms: 

 \begin{widetext}

\begin{eqnarray}
\mathscr{\sigma}^{\left(3\right)000}(t,t,t,t) & = & \frac{i\left(-\boldsymbol{e}^{4}\right)e^{-i\omega_{1}t}e^{-i\omega_{2}t}e^{-i\omega_{3}t}}{3!\omega_{1}\omega_{2}\omega_{3}}\int d\widetilde{\boldsymbol{k}}<|v^{abcd}|>_{0}\\
\mathscr{\sigma}_{abcd}^{\left(3\right)000}(\omega=\omega_{1}+\omega_{2}+\omega_{3}) & = & -\left(-\frac{\boldsymbol{e}^{4}}{\hbar^{3}}\right)\frac{\hbar^{3}}{3!\omega_{1}\omega_{2}\omega_{3}}\iint d\widetilde{\boldsymbol{k}}d\widetilde{E}f(E)Tr\left(v^{abcd}g_{E}^{r-a}\right)\label{eq:3000}\\
 & = & \left(-\frac{\boldsymbol{e}^{4}}{\hbar^{3}}\right)\frac{i\hbar^{3}}{3!\omega_{1}\omega_{2}\omega_{3}}\int d\widetilde{\boldsymbol{k}}\sum_{n}f(\varepsilon_{nk})v_{nn}^{abcd}\nonumber 
\end{eqnarray}

\begin{eqnarray}
\mathscr{\sigma}_{abcd}^{\left(3\right)100}(t,t,t,t^{\prime}) & = & \frac{\left(-\boldsymbol{e}^{4}\right)e^{-i\omega_{1}t}e^{-i\omega_{2}t}e^{-i\omega_{3}t^{\prime}}}{2\hbar\omega_{1}\omega_{2}\omega_{3}}\Theta_{t^{\prime}}^{t}\int d\widetilde{\boldsymbol{k}}<|\left[v^{abc}(t),v^{d}(t^{\prime})\right]|>_{0}
\end{eqnarray}

\begin{eqnarray}
 &  & \mathscr{\sigma}_{abcd}^{\left(3\right)100}(\omega=\omega_{1}+\omega_{2}+\omega_{3})\label{eq:3100}\\
 & = & -\left(-\frac{\boldsymbol{e}^{4}}{\hbar^{3}}\right)\frac{\hbar^{3}}{2\omega_{1}\omega_{2}\omega_{3}}\iint d\widetilde{\boldsymbol{k}}d\widetilde{E}f(E)Tr\left(v^{abc}g_{E+\hbar\omega_{3}}^{r}v^{d}g_{E}^{r-a}+g_{E}^{r-a}v^{d}g_{E-\hbar\omega_{3}}^{a}v^{abc}\right)\nonumber \\
 & = & \left(-\frac{\boldsymbol{e}^{4}}{\hbar^{3}}\right)\frac{i\hbar^{3}}{2\omega_{1}\omega_{2}\omega_{3}}\int d\widetilde{\boldsymbol{k}}\sum_{nm}f_{nm}\frac{v_{nm}^{abc}v_{mn}^{d}}{\varepsilon_{nk}-\varepsilon_{mk}+\hbar\omega_{3}+i\eta}\nonumber 
\end{eqnarray}
\begin{eqnarray}
\mathscr{\sigma}_{abcd}^{\left(3\right)200}(t,t,t^{\prime},t^{\prime}) & = & \frac{\left(-\boldsymbol{e}^{4}\right)e^{-i\omega_{1}t}e^{-i\omega_{2}t^{\prime}}e^{-i\omega_{3}t^{\prime}}}{2\hbar\omega_{1}\omega_{2}\omega_{3}}\Theta_{t^{\prime}}^{t}\int d\widetilde{\boldsymbol{k}}<|\left[v^{ab}(t),v^{cd}(t^{\prime})\right]|>_{0}
\end{eqnarray}

\begin{eqnarray}
 &  & \mathscr{\sigma}_{abcd}^{\left(3\right)200}(\omega=\omega_{1}+\omega_{2}+\omega_{3})\label{eq:3200}\\
 & = & -\left(-\frac{\boldsymbol{e}^{4}}{\hbar^{3}}\right)\frac{\hbar^{3}}{2\omega_{1}\omega_{2}\omega_{3}}\iint d\widetilde{\boldsymbol{k}}d\widetilde{E}f(E)Tr\left(v^{ab}g_{E+\hbar\omega_{2}+\hbar\omega_{3}}^{r}v^{cd}g_{E}^{r-a}+g_{E}^{r-a}v^{cd}g_{E-\hbar\omega_{2}-\hbar\omega_{3}}^{a}v^{ab}\right)\nonumber \\
 & = & \left(-\frac{\boldsymbol{e}^{4}}{\hbar^{3}}\right)\frac{i\hbar^{3}}{2\omega_{1}\omega_{2}\omega_{3}}d\widetilde{\boldsymbol{k}}\sum_{nm}f_{nm}\frac{v_{nm}^{ab}v_{mn}^{cd}}{\varepsilon_{nk}-\varepsilon_{mk}+\hbar\omega_{2}+\hbar\omega_{3}+i\eta}\nonumber 
\end{eqnarray}
 
\begin{eqnarray}
\mathscr{\sigma}_{abcd}^{\left(3\right)300}(t,t^{\prime},t^{\prime},t^{\prime}) & = & \frac{\left(-\boldsymbol{e}^{4}\right)e^{-i\omega_{1}t^{\prime}}e^{-i\omega_{2}t^{\prime}}e^{-i\omega_{3}t^{\prime}}}{3!\hbar\omega_{1}\omega_{2}\omega_{3}}\Theta_{t^{\prime}}^{t}\int d\widetilde{\boldsymbol{k}}<|\left[v^{a}(t),v^{bcd}(t^{\prime})\right]|>_{0}
\end{eqnarray}

\begin{eqnarray}
 &  & \mathscr{\sigma}_{abcd}^{\left(3\right)300}(\omega=\omega_{1}+\omega_{2}+\omega_{3})\label{eq:3300}\\
 & = & -\left(-\frac{\boldsymbol{e}^{4}}{\hbar^{3}}\right)\frac{\hbar^{3}}{3!\hbar\omega_{1}\omega_{2}\omega_{3}}\iint d\widetilde{\boldsymbol{k}}d\widetilde{E}f(E)Tr\left(v^{a}g_{E+\hbar\omega}^{r}v^{bcd}g_{E}^{r-a}+g_{E}^{r-a}v^{bcd}g_{E-\hbar\omega}^{a}v^{a}\right)\nonumber \\
 & = & \left(-\frac{\boldsymbol{e}^{4}}{\hbar^{3}}\right)\frac{i\hbar^{3}}{3!\hbar\omega_{1}\omega_{2}\omega_{3}}d\widetilde{\boldsymbol{k}}\sum_{nm}f_{nm}\frac{v_{nm}^{a}v_{mn}^{bcd}}{\varepsilon_{nk}-\varepsilon_{mk}+\hbar\omega+i\eta}\nonumber 
\end{eqnarray}
\begin{eqnarray}
\mathscr{\sigma}_{abcd}^{\left(3\right)110}(t,t,t^{\prime},t^{\prime\prime}) & = & -\frac{i\left(-\boldsymbol{e}^{4}\right)e^{-i\omega_{1}t}e^{-i\omega_{2}t^{\prime}}e^{-i\omega_{3}t^{\prime\prime}}}{\hbar^{2}\omega_{1}\omega_{2}\omega_{3}}\Theta_{t^{\prime}}^{t}\Theta_{t^{\prime\prime}}^{t^{\prime}}\int d\widetilde{\boldsymbol{k}}<|\left[\left[v^{ab}(t),v^{c}(t^{\prime})\right],v^{d}(t^{\prime\prime})\right]|>_{0}
\end{eqnarray}

\begin{eqnarray}
 &  & \mathscr{\sigma}_{abcd}^{\left(3\right)110}(\omega=\omega_{1}+\omega_{2}+\omega_{3})\label{eq:3110}\\
 & = & -\left(-\frac{\boldsymbol{e}^{4}}{\hbar^{3}}\right)\frac{i\hbar^{3}}{\omega_{1}\omega_{2}\omega_{3}}\iiint d\widetilde{\boldsymbol{k}}d\widetilde{E}d\widetilde{E}^{\prime}\left[f(E)-f(E^{\prime})\right]Tr\left(\frac{v^{ab}g_{E+\hbar\omega_{2}+\hbar\omega_{3}}^{r}v^{c}g_{E^{\prime}}^{r-a}\boldsymbol{J}_{c}g_{E}^{r-a}}{E+\hbar\omega_{3}-E^{\prime}+i\eta}+\frac{g_{E}^{r-a}v^{d}g_{E^{\prime}}^{r-a}v^{c}g_{E-\hbar\omega_{2}-\hbar\omega_{3}}^{a}v^{ab}}{E-\hbar\omega_{3}-E^{\prime}-i\eta}\right)\nonumber \\
 & = & -\left(-\frac{\boldsymbol{e}^{4}}{\hbar^{3}}\right)\frac{i\hbar^{3}}{\omega_{1}\omega_{2}\omega_{3}}\int d\widetilde{\boldsymbol{k}}\sum_{lmn}f_{nl}\left[\frac{v_{lm}^{ab}v_{mn}^{c}v_{nl}^{d}}{\left(\varepsilon_{lk}-\varepsilon_{nk}+\hbar\omega_{3}+i\eta\right)\left(\varepsilon_{lk}-\varepsilon_{mk}+\hbar\omega_{2}+\hbar\omega_{3}+i\eta\right)}\right.\nonumber \\
 &  & \,\,\,\,\,\,\,\,\,\,\,\,\,\,\,\,\,\,\,\,\,\,\,\,\,\,\,\,\,\,\,\,\,\,\,\,\,\,\,\,\,\,\,\,\,\left.+\frac{v_{ln}^{d}v_{nm}^{c}v_{ml}^{ab}}{\left(\varepsilon_{lk}-\varepsilon_{nk}-\hbar\omega_{3}-i\eta\right)\left(\varepsilon_{lk}-\varepsilon_{mk}-\hbar\omega_{2}-\hbar\omega_{3}-i\eta\right)}\right]\nonumber 
\end{eqnarray}
\begin{eqnarray}
\mathscr{\sigma}_{abcd}^{\left(3\right)210}(t,t^{\prime},t^{\prime},t^{\prime\prime}) & = & -\frac{i\left(-\boldsymbol{e}^{4}\right)e^{-i\omega_{1}t^{\prime}}e^{-i\omega_{2}t^{\prime}}e^{-i\omega_{3}t^{\prime\prime}}}{2\hbar^{2}\omega_{1}\omega_{2}\omega_{3}}\Theta_{t^{\prime}}^{t}\Theta_{t^{\prime\prime}}^{t^{\prime}}\int d\widetilde{\boldsymbol{k}}<|\left[\left[v^{a}(t),v^{bc}(t^{\prime})\right],v^{d}(t^{\prime\prime})\right]|>_{0}
\end{eqnarray}

\begin{eqnarray}
 &  & \mathscr{\sigma}_{abcd}^{3210}(\omega=\omega_{1}+\omega_{2}+\omega_{3})\label{eq:3210}\\
 & = & -\left(-\frac{\boldsymbol{e}^{4}}{\hbar^{3}}\right)\frac{i\hbar^{3}}{2\omega_{1}\omega_{2}\omega_{3}}\iiint d\widetilde{\boldsymbol{k}}d\widetilde{E}d\widetilde{E}^{\prime}\left[f(E)-f(E^{\prime})\right]Tr\left(\frac{v^{a}g_{E+\hbar\omega}^{r}v^{bc}g_{E^{\prime}}^{r-a}v^{d}g_{E}^{r-a}}{E+\hbar\omega_{3}-E^{\prime}+i\eta}+\frac{g_{E}^{r-a}v^{d}g_{E^{\prime}}^{r-a}v^{bc}g_{E-\hbar\omega}^{a}v^{a}}{E-\hbar\omega_{3}-E^{\prime}-i\eta}\right)\nonumber \\
 & = & -\left(-\frac{\boldsymbol{e}^{4}}{\hbar^{3}}\right)\frac{i\hbar^{3}}{2\omega_{1}\omega_{2}\omega_{3}}\int d\widetilde{\boldsymbol{k}}\sum_{lmn}f_{nl}\left[\frac{v_{lm}^{a}v_{mn}^{bc}v_{nl}^{d}}{\left(\varepsilon_{lk}-\varepsilon_{nk}+\hbar\omega_{3}+i\eta\right)\left(\varepsilon_{lk}-\varepsilon_{mk}+\hbar\omega+i\eta\right)}\right.\nonumber \\
 &  & \,\,\,\,\,\,\,\,\,\,\,\,\,\,\,\,\,\,\,\,\,\,\,\,\,\,\,\,\,\,\,\,\,\,\,\,\,\,\,\,\,\,\,\,\,\left.+\frac{v_{ln}^{d}v_{nm}^{bc}v_{ml}^{a}}{\left(\varepsilon_{lk}-\varepsilon_{nk}-\hbar\omega_{3}-i\eta\right)\left(\varepsilon_{lk}-\varepsilon_{mk}-\hbar\omega-i\eta\right)}\right]\nonumber 
\end{eqnarray}
\begin{eqnarray}
\mathscr{\sigma}_{abcd}^{\left(3\right)120}(t,t^{\prime},t^{\prime\prime},t^{\prime\prime}) & = & -\frac{i\left(-\boldsymbol{e}^{4}\right)e^{-i\omega_{1}t^{\prime}}e^{-i\omega_{2}t^{\prime\prime}}e^{-i\omega_{3}t^{\prime\prime}}}{2\hbar^{2}\omega_{1}\omega_{2}\omega_{3}}\Theta_{t^{\prime}}^{t}\Theta_{t^{\prime\prime}}^{t^{\prime}}\int d\widetilde{\boldsymbol{k}}<|\left[\left[v^{a}(t),v^{b}(t^{\prime})\right],v^{cd}(t^{\prime\prime})\right]|>_{0}
\end{eqnarray}

\begin{eqnarray}
 &  & \mathscr{\sigma}_{abcd}^{3120}(\omega=\omega_{1}+\omega_{2}+\omega_{3})\label{eq:3120}\\
 & = & -\left(-\frac{\boldsymbol{e}^{4}}{\hbar^{3}}\right)\frac{i\hbar^{3}}{2\omega_{1}\omega_{2}\omega_{3}}\iiint d\widetilde{\boldsymbol{k}}d\widetilde{E}d\widetilde{E}^{\prime}\left[f(E)-f(E^{\prime})\right]Tr\left(\frac{v^{a}g_{E+\hbar\omega}^{r}v^{b}g_{E^{\prime}}^{r-a}v^{cd}g_{E}^{r-a}}{E+\hbar\omega_{2}+\hbar\omega_{3}-E^{\prime}+i\eta}+\frac{g_{E}^{r-a}v^{cd}g_{E^{\prime}}^{r-a}v^{b}g_{E-\hbar\omega}^{a}v^{a}}{E-\hbar\omega_{2}-\hbar\omega_{3}-E^{\prime}-i\eta}\right)\nonumber \\
 & = & -\left(-\frac{\boldsymbol{e}^{4}}{\hbar^{3}}\right)\frac{i\hbar^{3}}{2\omega_{1}\omega_{2}\omega_{3}}\int d\widetilde{\boldsymbol{k}}\sum_{lmn}f_{nl}\left[\frac{v_{lm}^{a}v_{mn}^{b}v_{nl}^{cd}}{\left(\varepsilon_{lk}-\varepsilon_{nk}+\hbar\omega_{2}+\hbar\omega_{3}+i\eta\right)\left(\varepsilon_{lk}-\varepsilon_{mk}+\hbar\omega+i\eta\right)}\right.\nonumber \\
 &  & \,\,\,\,\,\,\,\,\,\,\,\,\,\,\,\,\,\,\,\,\,\,\,\,\,\,\,\,\,\,\,\,\,\,\,\,\,\,\,\,\,\,\,\,\,\left.+\frac{v_{ln}^{cd}v_{nm}^{b}v_{ml}^{a}}{\left(\varepsilon_{lk}-\varepsilon_{nk}-\hbar\omega_{2}-\hbar\omega_{3}-i\eta\right)\left(\varepsilon_{lk}-\varepsilon_{mk}-\hbar\omega-i\eta\right)}\right]\nonumber 
\end{eqnarray}

\begin{eqnarray}
\mathscr{\sigma}_{abcd}^{\left(3\right)111}(t,t^{\prime},t^{\prime\prime},t^{\prime\prime\prime}) & = & -\frac{\left(-\boldsymbol{e}^{4}\right)e^{-i\omega_{1}t^{\prime}}e^{-i\omega_{2}t^{\prime\prime}}e^{-i\omega_{3}t^{\prime\prime\prime}}}{\hbar^{3}\omega_{1}\omega_{2}\omega_{3}}\Theta_{t^{\prime}}^{t}\Theta_{t^{\prime\prime}}^{t^{\prime}}\Theta_{t^{\prime\prime\prime}}^{t^{\prime\prime}}\int d\widetilde{\boldsymbol{k}}<|\left[\left[\left[v^{a}(t),v^{b}(t^{\prime})\right],v^{c}(t^{\prime\prime})\right],v^{d}(t^{\prime\prime\prime})\right]|>_{0}
\end{eqnarray}

\begin{eqnarray}
 &  & \mathscr{\sigma}_{abcd}^{\left(3\right)111}(\omega=\omega_{1}+\omega_{2}+\omega_{3})\label{eq:3111}\\
 & = & +\left(-\frac{\boldsymbol{e}^{4}}{\hbar^{3}}\right)\frac{i\hbar^{3}}{\omega_{1}\omega_{2}\omega_{3}}\iiint d\widetilde{\boldsymbol{k}}d\widetilde{E}d\widetilde{E}^{\prime}\left[f(E)-f(E^{\prime})\right]\times\nonumber \\
 &  & \,\,\,\,\,\,\,\,\,\,\,\,\,\,\,\,\,\,\,\,\,\,\,\,\,\,\,\,\,\,\,\,\,\,\,\,\,\,\,\,\,\,\,\,\,Tr\left(\frac{v^{a}g_{E+\hbar\omega}^{r}v^{b}g_{E+\hbar\omega_{2}+\hbar\omega_{3}}^{r}v^{c}g_{E^{\prime}}^{r-a}v^{d}g_{E}^{r-a}}{E+\hbar\omega_{3}-E^{\prime}+i\eta}+\frac{g_{E}^{r-a}v^{d}g_{E^{\prime}}^{r-a}v^{c}g_{E-\hbar\omega_{2}-\hbar\omega_{3}}^{a}v^{b}g_{E-\hbar\omega}^{a}v^{a}}{E-\hbar\omega_{3}-E^{\prime}-i\eta}\right)\nonumber \\
 &  & +\left(-\frac{\boldsymbol{e}^{4}}{\hbar^{3}}\right)\frac{\hbar^{3}}{\omega_{1}\omega_{2}\omega_{3}}\iiiint d\widetilde{\boldsymbol{k}}d\widetilde{E}d\widetilde{E}^{\prime}d\widetilde{E}^{\prime\prime}\left[f(E^{\prime})-f(E^{\prime\prime})\right]\times\nonumber \\
 &  & \,\,\,Tr\left[\frac{v^{a}g_{E+\hbar\omega}^{r}v^{b}g_{E^{\prime}}^{r-a}v^{d}g_{E^{\prime\prime}}^{r-a}v^{c}g_{E}^{r-a}}{\left(E^{\prime}-\hbar\omega_{3}-E^{\prime\prime}-i\eta\right)\left(E+\hbar\omega_{2}+\hbar\omega_{3}-E^{\prime}+i\eta\right)}+\frac{g_{E}^{r-a}v^{c}g_{E^{\prime\prime}}^{r-a}v^{d}g_{E^{\prime}}^{r-a}v^{b}g_{E-\hbar\omega}^{a}v^{a}}{\left(E^{\prime}+\hbar\omega_{3}-E^{\prime\prime}+i\eta\right)\left(E-\hbar\omega_{2}-\hbar\omega_{3}-E^{\prime}-i\eta\right)}\right]\nonumber \\
 & = & \left(-\frac{\boldsymbol{e}^{4}}{\hbar^{3}}\right)\frac{i\hbar^{3}}{\omega_{1}\omega_{2}\omega_{3}}\int d\widetilde{\boldsymbol{k}}\sum_{lmnp}\left[f_{pl}\frac{v_{lm}^{a}v_{mn}^{b}v_{np}^{c}v_{pl}^{d}}{\left(\varepsilon_{lk}+\hbar\omega_{3}-\varepsilon_{pk}+i\eta\right)\left(\varepsilon_{lk}+\hbar\omega_{2}+\hbar\omega_{3}-\varepsilon_{nk}+i\eta\right)\left(\varepsilon_{lk}+\hbar\omega-\varepsilon_{mk}+i\eta\right)}\right.\nonumber \\
 &  & \,\,\,\,\,\,\,\,\,\,\,\,\,\,\,\,\,\,\,\,\,\,\,\,\,\,\,\,\,\,\,\,\,\,\,\,+f_{pl}\frac{v_{lp}^{d}v_{pn}^{c}v_{nm}^{b}v_{ml}^{a}}{\left(\varepsilon_{lk}-\hbar\omega_{3}-\varepsilon_{pk}-i\eta\right)\left(\varepsilon_{lk}-\hbar\omega_{2}-\hbar\omega_{3}-\varepsilon_{nk}-i\eta\right)\left(\varepsilon_{lk}-\hbar\omega-\varepsilon_{mk}-i\eta\right)}\nonumber \\
 &  & \,\,\,\,\,\,\,\,\,\,\,\,\,\,\,\,\,\,\,\,\,\,\,\,\,\,\,\,\,\,\,\,\,\,\,\,+f_{pn}\frac{v_{lm}^{a}v_{mn}^{b}v_{np}^{d}v_{pl}^{c}}{\left(\varepsilon_{pk}+\hbar\omega_{3}-\varepsilon_{nk}+i\eta\right)\left(\varepsilon_{lk}+\hbar\omega_{2}+\hbar\omega_{3}-\varepsilon_{nk}+i\eta\right)\left(\varepsilon_{lk}+\hbar\omega-\varepsilon_{mk}+i\eta\right)}\nonumber \\
 &  & \,\,\,\,\,\,\,\,\,\,\,\,\,\,\,\,\,\,\,\,\,\,\,\,\,\,\,\,\,\,\,\,\,\,\,\,+\left.f_{pn}\frac{v_{lp}^{c}v_{pn}^{d}v_{nm}^{b}v_{ml}^{a}}{\left(\varepsilon_{pk}-\hbar\omega_{3}-\varepsilon_{nk}-i\eta\right)\left(\varepsilon_{lk}-\hbar\omega_{2}-\hbar\omega_{3}-\varepsilon_{nk}-i\eta\right)\left(\varepsilon_{lk}-\hbar\omega-\varepsilon_{mk}-i\eta\right)}\right]\nonumber \\
\nonumber 
\end{eqnarray}

(seeing Appendix \ref{sec:Aa}\&\ref{sec:Ab} for details).

 \end{widetext}

Recently, third-order nonlinear Hall effect driven by two photons
and DC electric field is observed in experiments\citep{ma2019nonlinear}.
In our method, we can derive the formula for third-order Hall effect:

\begin{eqnarray}
 &  & \bar{\sigma}_{NHE}^{\left(3\right)}\left(\omega\right)\label{eq:3NHE}\\
 & = & \underset{\alpha}{\sum}\left[\mathscr{\sigma}_{abcd}^{\left(3\right)\alpha}(\pm\omega,\mp\omega,\omega_{0})+\mathscr{\sigma}_{abcd}^{\left(3\right)\alpha}(\pm\omega,\mp\omega,-\omega_{0})\right.\nonumber \\
 &  & \,\,\,+\mathscr{\sigma}_{abcd}^{\left(3\right)\alpha}(\pm\omega,\omega_{0},\mp\omega)+\mathscr{\sigma}_{abcd}^{\left(3\right)\alpha}(\mp\omega,-\omega_{0},\pm\omega)\nonumber \\
 &  & \left.+\mathscr{\sigma}_{abcd}^{\left(3\right)\alpha}(\omega_{0},\pm\omega,\mp\omega)+\mathscr{\sigma}_{abcd}^{\left(3\right)\alpha}(-\omega_{0},\pm\omega,\mp\omega)\right]{}_{\omega_{0}\rightarrow0}\nonumber \\
\nonumber 
\end{eqnarray}
, where for the limit of space, we do not list the result of third-order
nonlinear Hall effect in Green's function. Comparing with Ref. \citep{ma2019nonlinear},
their method is based on second order effect (shift current and injection
current) and modifies the Fermi-Dirac distribution under the affect
of DC electric field, which only involved in two-band terms. Combining
with Eq. \ref{eq:3000}-\ref{eq:3111} and Eq. \ref{eq:3NHE}, we
can see that in our formulas, there are not only involved in intraband
terms and two-band terms, but also three-band and four-band terms,
which implies the results of Eq. \ref{eq:3NHE} are more comprehensive
to describe general situations for third-order nonlinear Hall effect.

For optical current researches, most papers are talking about terms
like $\mathscr{\sigma}_{ab}^{\left(1\right)1}$, $\mathscr{\sigma}_{abc}^{\left(2\right)11}$,
$\mathscr{\sigma}_{abcd}^{\left(3\right)111}$and ignore others. One
reason may be the models they used are generally $k\cdot p$ models,
which usually include linear terms of $k$ only in Hamiltonian, and
$v^{bc\cdots}=\frac{\partial^{n}H_{k}}{\hbar^{n}\partial k^{b}\partial k^{c}\cdots}$
vanishes in high orders. To be cleared that there is no evidence that
these terms dominant while the others are negligible. For $\mathscr{\sigma}_{ab}^{\left(1\right)0}$,
$\mathscr{\sigma}_{abc}^{\left(2\right)00}$, $\mathscr{\sigma}_{abcd}^{\left(3\right)000}\cdots$,
it is obvious that they are single-band terms, they shares similar
forms in Green's function formulas, and also in Feynman Diagrams (Fig.\ref{fig1},
\ref{fig2}, \ref{fig3}). And also, $\mathscr{\sigma}_{ab}^{\left(1\right)1}$,
$\mathscr{\sigma}_{abc}^{\left(2\right)10}$, $\mathscr{\sigma}_{abc}^{\left(2\right)20}$,
$\mathscr{\sigma}_{abcd}^{\left(3\right)100}$, $\mathscr{\sigma}_{abcd}^{\left(3\right)200}$,
$\mathscr{\sigma}_{abcd}^{\left(3\right)300}\cdots$shares similar
forms involving two bands, while $\mathscr{\sigma}_{abc}^{\left(2\right)11}$,
$\mathscr{\sigma}_{abcd}^{\left(3\right)110}$, $\mathscr{\sigma}_{abcd}^{\left(3\right)210}$,
$\mathscr{\sigma}_{abcd}^{\left(3\right)120}\cdots$ involve three
bands. Furthermore, we can image that $\mathscr{\sigma}_{abcd}^{\left(3\right)111}$,
$\mathscr{\sigma}_{abcde}^{\left(4\right)1110}$, $\mathscr{\sigma}_{abcde}^{\left(4\right)2110}$,
$\mathscr{\sigma}_{abcde}^{\left(4\right)1210}$, $\mathscr{\sigma}_{abcde}^{\left(4\right)1120}\cdots$
have interband form with four bands. All these terms in Lehmann representation
are comparable in with diagrammatic approach in Ref. \citep{prb2019-JMoore-diagrammatic-formula}.
Using similar method of density operator, Wick's theorem, Langreth
rule and Fourier transform, higher-order responses could be also derived
if someone needs. 

\section{Discussion for many-body effect}

\begin{figure}
\begin{centering}
\includegraphics[width=0.6\columnwidth]{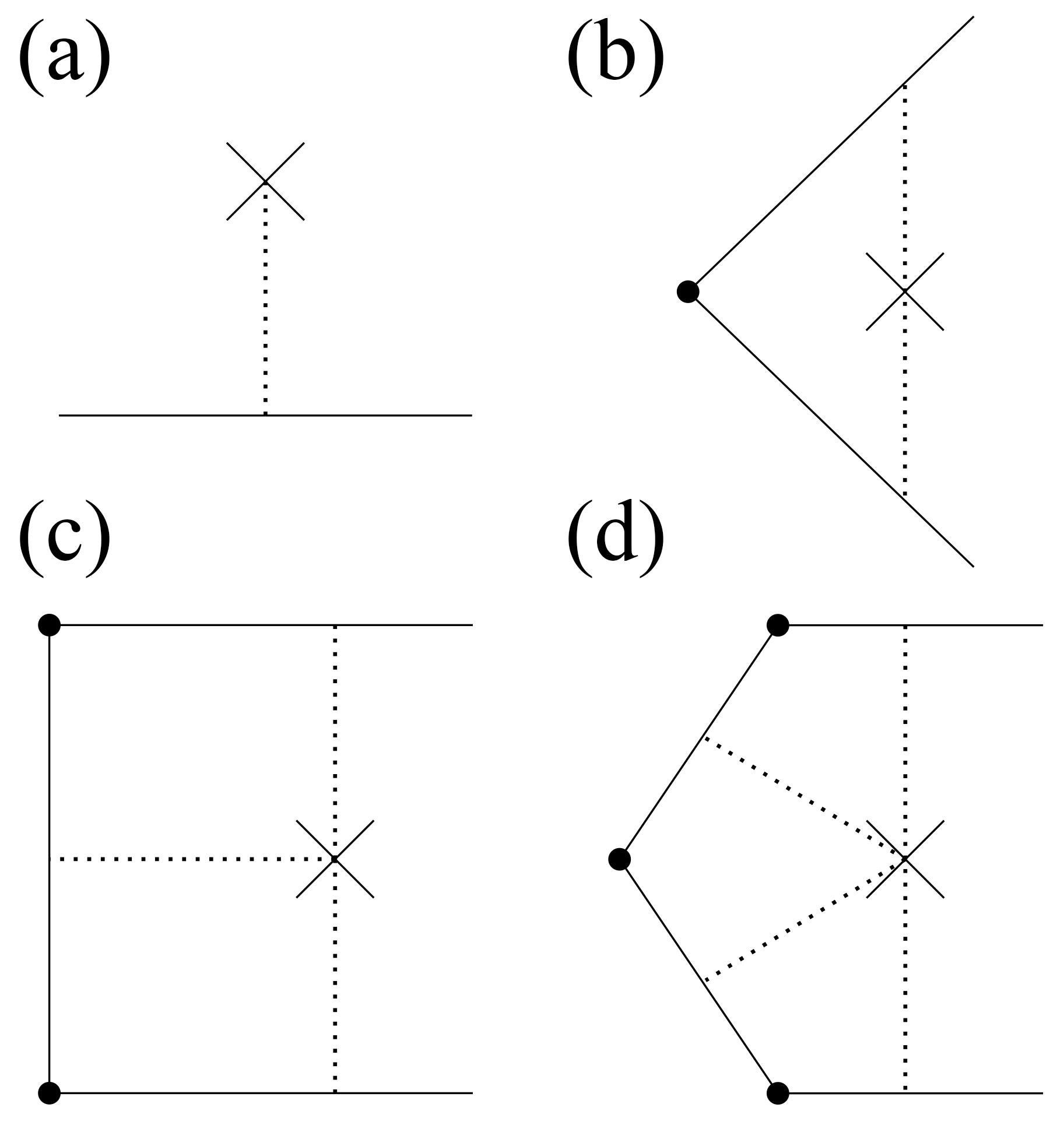} 
\par\end{centering}
\caption{\textbf{Feynman diagrams for the correction of disorder effect.} (a)
First-order correction for single-particle Green's function. (b) Second-order
one-vertex correction. (c) Third-order double-vertex correction. (d)
Fourth-order triple-vertex correction.\label{fig4} }
\end{figure}

Generally, Green's function formula is a convenient way to consider
many-body effect in excitation-response process, with some methods
like self energy correction, vertex correction in disorder effect,
or random phase approximation ( RPA ) method in considering strong-correlation
or electron-phonon coupling. 

Here we take disorder effect as an example with disorder Hamiltonian
$H=H_{0}+V_{D}$ , where $V_{D}=\sum_{i}p_{i}V_{D^{i}}$ is the perturbation
of disorder with possibility distribution $p_{i},\,\sum_{i}p_{i}=1$.
At beginning, in first-order correction, we correct Green's function
$\tilde{g}_{E}^{(1)}=\sum_{i}p_{i}g_{E}V_{D^{i}}g_{E}$ ( Fig.\ref{fig4}a),
where $g_{E}$ here represents retarded or advanced Green's function. 

Then in second-order correction, except for Green's function $\tilde{g}_{E}^{(2)}=\sum_{i,j\neq i}p_{i}p_{j}g_{E}V_{D^{i}}g_{E}V_{D^{j}}g_{E}$,
we have correction for each vertex operator like $v^{\alpha}$ :

\begin{eqnarray}
\widetilde{v}_{(2)}^{\alpha} & = & \sum_{i}p_{i}^{2}V_{D^{i}}g_{E}^{r/a}v^{\alpha}g_{E^{\prime}}^{r/a}V_{D^{i}}\label{eq:v-corr}
\end{eqnarray}
, which is second-order single-vertex correction ( Fig.\ref{fig4}b).
We can also consider second-order double-vertex correction $\widetilde{v^{\alpha}g_{E}v^{\beta}}_{(2)}$:
\begin{eqnarray}
\widetilde{v^{\alpha}g_{E}v^{\beta}}_{(2)} & = & \sum_{i}p_{i}^{2}V_{D^{i}}g_{E^{\prime}}^{r/a}v^{\alpha}g_{E}v^{\beta}g_{E^{\prime\prime}}^{r/a}V_{D^{i}}
\end{eqnarray}
. Besides, we have higher-order correction with more vertexes, such
as third-order double-vertex correction ( Fig.\ref{fig4}c ) and fourth-order
triple-vertex correction ( Fig.\ref{fig4}d ).

By summing up all the cases, the response function of the correction
with the order we want could be got. For example, the second-order
correction for the term $v^{a}g_{E+\hbar\omega}^{r}v^{b}g_{E^{\prime}}^{r-a,(2)}v^{c}g_{E}^{r-a}$
in second-order response $\mathscr{\sigma}_{abc}^{\left(2\right)11}(\omega_{1},\omega_{2})$
is 

\begin{eqnarray}
 &  & Tr\left(\overline{v^{a}g_{E+\hbar\omega}^{r}v^{b}g_{E^{\prime}}^{r-a,(2)}v^{c}g_{E}^{r-a}}_{(2)}\right)\\
 & = & Tr\left(\frac{v^{a}\tilde{g}_{E+\hbar\omega}^{r,(2)}v^{b}g_{E^{\prime}}^{r-a}v^{c}g_{E}^{r-a}}{E+\hbar\omega_{2}-E^{\prime}+i\eta}+\right.\nonumber \\
 &  & \,\,\,\,\frac{v^{a}g_{E+\hbar\omega}^{r}v^{b}\tilde{g}_{E^{\prime}}^{r-a,(2)}v^{c}g_{E}^{r-a}}{E+\hbar\omega_{2}-E^{\prime}+i\eta}+\nonumber \\
 &  & \,\,\,\,\frac{v^{a}g_{E+\hbar\omega}^{r}v^{b}g_{E^{\prime}}^{r-a}v^{c}\tilde{g}_{E}^{r-a,(2)}}{E+\hbar\omega_{2}-E^{\prime}+i\eta}+\nonumber \\
 &  & \,\,\,\,\frac{v^{a}\tilde{g}_{E+\hbar\omega}^{r,(1)}v^{b}g_{E^{\prime}}^{r-a,(1)}v^{c}g_{E}^{r-a}}{E+\hbar\omega_{2}-E^{\prime}+i\eta}+\nonumber \\
 &  & \,\,\,\,\frac{v^{a}g_{E+\hbar\omega}^{r}v^{b}\tilde{g}_{E^{\prime}}^{r-a,(1)}v^{c}\tilde{g}_{E}^{r-a,(1)}}{E+\hbar\omega_{2}-E^{\prime}+i\eta}+\nonumber \\
 &  & \,\,\,\,\frac{v^{a}\tilde{g}_{E+\hbar\omega}^{r,(1)}v^{b}g_{E^{\prime}}^{r-a}v^{c}\tilde{g}_{E}^{r-a,(1)}}{E+\hbar\omega_{2}-E^{\prime}+i\eta}+\nonumber \\
 &  & \,\,\,\,\frac{\tilde{v}_{(2)}^{a}g_{E+\hbar\omega}^{r}v^{b}g_{E^{\prime}}^{r-a}v^{c}g_{E}^{r-a}}{E+\hbar\omega_{2}-E^{\prime}+i\eta}+\nonumber \\
 &  & \,\,\,\,\frac{v^{a}g_{E+\hbar\omega}^{r}\tilde{v}_{(2)}^{b}g_{E^{\prime}}^{r-a}v^{c}g_{E}^{r-a}}{E+\hbar\omega_{2}-E^{\prime}+i\eta}+\nonumber \\
 &  & \,\,\,\,\left.\frac{v^{a}g_{E+\hbar\omega}^{r}v^{b}g_{E^{\prime}}^{r-a}\tilde{v}_{(2)}^{c}g_{E}^{r-a}}{E+\hbar\omega_{2}-E^{\prime}+i\eta}-D.C.\right)\nonumber 
\end{eqnarray}
, where $D.C.$ means double counting terms. According to coherent
potential approximation ( CPA )\citep{levin1970electronic,zhou2017general},
corrected Green's function $\tilde{g}_{E}^{(i)}$ related terms and
$D.C.$ could be eliminated, which simplifies formulas further. To
be cleared that there is no second-order double-vertex correction
term, because it is the same as second-order single-vertex correction
term in $\mathscr{\sigma}_{abc}^{\left(2\right)11}(\omega_{1},\omega_{2})$.
We take an example of a model with disorder in App.\ref{sec:Ad} .

\section{Summary}

In a conclusion, using density operator method, we derive the equations
of linear and nonlinear optical responses in Green's function formulas.
The form of novel physical quantities in Green's function formulas
are also got by linear combination of linear or nonlinear optical
conductivities, which pave a road to study higher-order responses.
Taking advantages of Green's function, one can consider many-body
effects in nonlinear responses based on the methods of self energy
correction, vertex correction or RPA in different cases.

\section{Acknowledgments}

The authors thank Jin Cao and Chenyi Zhou for useful discussion. 

 \begin{widetext}

\appendix

\section{Langreth rule for more than two particles \label{sec:Aa}}

In Langreth rule, to solve two-particle retarded green function 
\begin{eqnarray}
C^{r}(z_{1},z_{2}) & \equiv & \Theta(z_{1}-z_{2})\left[C^{>}(z_{1},z_{2})-C^{<}(z_{1},z_{2})\right]\nonumber \\
 & \equiv & \left(-i\right)^{2}\Theta(z_{1}-z_{2})\left\langle \left[O\left(z_{1}\right),O\left(z_{2}\right)\right]\right\rangle \nonumber \\
C^{>}(z_{1},z_{2}) & \equiv & \left(-i\right)^{2}\left\langle O\left(z_{1}\right)O\left(z_{2}\right)\right\rangle ,O\left(z_{1}\right)\equiv\underset{ij}{\sum}c_{i}^{\dagger}(z_{1})c_{j}(z_{1})\nonumber \\
C^{<}(z_{1},z_{2}) & \equiv & \left(-i\right)^{2}\left\langle O\left(z_{2}\right)O\left(z_{1}\right)\right\rangle \nonumber \\
C^{T}(z_{1},z_{2}) & \equiv & \left(-i\right)^{2}\left\langle T\left[O\left(z_{1}\right)O\left(z_{2}\right)\right]\right\rangle \nonumber \\
 & = & \left(-i\right)^{2}\Theta(z_{1}-z_{2})\left\langle O\left(z_{1}\right)O\left(z_{2}\right)\right\rangle +\left(-i\right)^{2}\Theta(z_{2}-z_{1})\left\langle O\left(z_{2}\right)O\left(z_{1}\right)\right\rangle \nonumber \\
 & = & \Theta(z_{1}-z_{2})C^{>}(z_{1},z_{2})+\Theta(z_{2}-z_{1})C^{<}(z_{1},z_{2})\label{eq:ct}\\
C^{\widetilde{T}}(z_{1},z_{2}) & \equiv & \Theta(z_{2}-z_{1})C^{>}(z_{1},z_{2})+\Theta(z_{1}-z_{2})C^{<}(z_{1},z_{2})\label{eq:cat}
\end{eqnarray}
Here, $C^{r/>/</T}$ is two-particle retarded / greater / lesser /
time-order green function, we have follow solution:
\begin{eqnarray}
C^{T}(z_{1},z_{2}) & = & \left(-i\right)^{2}\left\langle T\left[O\left(z_{1}\right)O\left(z_{2}\right)\right]\right\rangle \nonumber \\
 & = & \left(-i\right)^{2}\underset{abcd}{\sum}\left\langle T\left[c_{a}^{\dagger}(z_{1})c_{b}(z_{1})c_{c}^{\dagger}(z_{2})c_{d}(z_{2})\right]\right\rangle \nonumber \\
 & = & -\left(-i\right)^{2}\underset{abcd}{\sum}\left\langle T\left[c_{d}(z_{2})c_{a}^{\dagger}(z_{1})c_{b}(z_{1})c_{c}^{\dagger}(z_{2})\right]\right\rangle \nonumber \\
 & = & -\left(-i\right)^{2}\underset{ad}{\sum}\left\langle T\left[c_{d}(z_{2})c_{a}^{\dagger}(z_{1})\right]\right\rangle \underset{bc}{\sum}\left\langle T\left[c_{b}(z_{1})c_{c}^{\dagger}(z_{2})\right]\right\rangle \nonumber \\
 & = & -a^{T}(z_{1},z_{2})b^{T}(z_{2},z_{1})\nonumber \\
 & = & -\left[\Theta(z_{1}-z_{2})a^{>}(z_{1},z_{2})+\Theta(z_{2}-z_{1})a^{<}(z_{1},z_{2})\right]\left[\Theta(z_{2}-z_{1})b^{>}(z_{2},z_{1})+\Theta(z_{1}-z_{2})b^{<}(z_{2},z_{1})\right]\label{eq:ct1}\\
\nonumber \\C^{\widetilde{T}}(z_{1},z_{2}) & = & -a^{\widetilde{T}}(z_{1},z_{2})b^{\widetilde{T}}(z_{2},z_{1})\nonumber \\
 & = & -\left[\Theta(z_{2}-z_{1})a^{>}(z_{1},z_{2})+\Theta(z_{1}-z_{2})a^{<}(z_{1},z_{2})\right]\left[\Theta(z_{1}-z_{2})b^{>}(z_{2},z_{1})+\Theta(z_{2}-z_{1})b^{<}(z_{2},z_{1})\right]\label{eq:ct2}\\
\nonumber 
\end{eqnarray}
,which is the result of Wick's theorem, where $G^{T(\widetilde{T})}$
is (anti-)time-ordered green function and $a/b$ is one partical green
function. 

So conbining Eq. \ref{eq:ct} \& \ref{eq:ct1} (Eq. \ref{eq:cat}
\& \ref{eq:ct2}), we can get $C^{>}(z_{1},z_{2})=-a^{>}(z_{1},z_{2})b^{<}(z_{2},z_{1})$
($C^{<}(z_{1},z_{2})=-a^{<}(z_{1},z_{2})b^{>}(z_{2},z_{1})$), and
further solve $C^{r}$:
\begin{eqnarray*}
C^{r}(z_{1},z_{2}) & = & \Theta(z_{1}-z_{2})\left[C^{>}(z_{1},z_{2})-C^{<}(z_{1},z_{2})\right]\\
 & = & -\Theta(z_{1}-z_{2})\left[a^{>}(z_{1},z_{2})b^{<}(z_{2},z_{1})-a^{<}(z_{1},z_{2})b^{>}(z_{2},z_{1})\right]\\
 & = & -\Theta(z_{1}-z_{2})\left[a^{>}(z_{1},z_{2})b^{<}(z_{2},z_{1})-a^{<}(z_{1},z_{2})b^{<}(z_{2},z_{1})+a^{<}(z_{1},z_{2})b^{<}(z_{2},z_{1})-a^{<}(z_{1},z_{2})b^{>}(z_{2},z_{1})\right]\\
 & = & -a^{r}(z_{1},z_{2})b^{<}(z_{2},z_{1})-a^{<}(z_{1},z_{2})b^{a}(z_{2},z_{1})\\
\\(i)^{2}C^{r}(z_{1},z_{2}) & = & \Theta(z_{1}-z_{2})\left\langle \left[O\left(z_{1}\right),O\left(z_{2}\right)\right]\right\rangle =a^{r}(z_{1},z_{2})b^{<}(z_{2},z_{1})+a^{<}(z_{1},z_{2})b^{a}(z_{2},z_{1})
\end{eqnarray*}

Similar to the method to solve two-particle retarded green function,
we can solve three-particle and four-particle retarded green function:

\begin{eqnarray*}
(i)^{3}D^{r}(z_{1},z_{2},z_{3}) & = & \Theta(z_{1}-z_{2})\Theta(z_{2}-z_{3})<|\left[\left[O\left(z_{1}\right),O\left(z_{2}\right)\right],O\left(z_{3}\right)\right]|>\\
 & = & +ia^{r}(z_{1},z_{2})b^{r}(z_{2},z_{3})c^{<}(z_{3},z_{1})+i\Theta_{23}a^{r}(z_{1},z_{2})b^{<}(z_{2},z_{3})\left[c^{<}(z_{3},z_{1})-c^{>}(z_{3},z_{1})\right]\\
 &  & +ia^{a}(z_{2},z_{1})b^{a}(z_{3},z_{2})c^{<}(z_{1},z_{3})+i\Theta_{23}a^{a}(z_{2},z_{1})b^{<}(z_{3},z_{2})\left[c^{>}(z_{1},z_{3})-c^{<}(z_{1},z_{3})\right]\\
\end{eqnarray*}

\begin{eqnarray*}
(i)^{4}F^{r}(z_{1},z_{2},z_{3},z_{4}) & = & \Theta(z_{1}-z_{2})\Theta(z_{2}-z_{3})\Theta(z_{3}-z_{4})<|\left[\left[\left[O\left(z_{1}\right),O\left(z_{2}\right)\right],O\left(z_{3}\right)\right],O\left(z_{4}\right)\right]|>\\
 & = & g^{r}(z_{1},z_{2})g^{r}(z_{2},z_{3})g^{r}(z_{3},z_{4})g^{<}(z_{4},z_{1})\\
 &  & +\Theta^{34}g^{r}(z_{1},z_{2})g^{r}(z_{2},z_{3})g^{<}(z_{3},z_{4})\left[g^{<}(z_{4},z_{1})-g^{>}(z_{4},z_{1})\right]\\
 &  & +\Theta^{23}g^{r}(z_{1},z_{2})g^{<}(z_{2},z_{4})g^{a}(z_{4},z_{3})\left[g^{<}(z_{3},z_{1})-g^{>}(z_{3},z_{1})\right]\\
 &  & +\Theta^{23}\Theta^{34}g^{r}(z_{1},z_{2})\left[g^{>}(z_{2},z_{4})-g^{<}(z_{2},z_{4})\right]g^{<}(z_{4},z_{3})\left[g^{<}(z_{3},z_{1})-g^{>}(z_{3},z_{1})\right]\\
 &  & +\Theta^{23}\Theta^{34}g^{a}(z_{2},z_{1})\left[g^{<}(z_{4},z_{2})-g^{>}(z_{4},z_{2})\right]g^{<}(z_{3},z_{4})\left[g^{>}(z_{1},z_{3})-g^{<}(z_{1},z_{3})\right]\\
 &  & +\Theta^{23}g^{a}(z_{2},z_{1})g^{<}(z_{4},z_{2})g^{r}(z_{3},z_{4})\left[g^{>}(z_{1},z_{3})-g^{<}(z_{1},z_{3})\right]\\
 &  & +\Theta^{34}g^{a}(z_{2},z_{1})g^{a}(z_{3},z_{2})g^{<}(z_{4},z_{3})\left[g^{>}(z_{1},z_{4})-g^{<}(z_{1},z_{4})\right]\\
 &  & +g^{a}(z_{2},z_{1})g^{a}(z_{3},z_{2})g^{a}(z_{4},z_{3})g^{<}(z_{1},z_{4})
\end{eqnarray*}
.

\section{Derivation details for Correlation Functions \label{sec:Ab}}

\begin{eqnarray*}
\mathscr{\sigma}_{ab}^{\left(1\right)0}(t,t) & = & \int d\widetilde{\boldsymbol{k}}<|\frac{\partial\hat{R}^{a}(t)}{\partial\mathbf{E}^{b}(t)}|>_{0}\\
 & = & \frac{\left(-\boldsymbol{e}^{2}\right)e^{-i\omega_{1}t}}{i\omega_{1}}\int d\widetilde{\boldsymbol{k}}<|v^{ab}|>_{0}\\
\mathscr{\sigma}_{ab}^{\left(1\right)0}(\omega=\omega_{1}) & = & \frac{1}{2\pi}\int_{-\infty}^{+\infty}dte^{i\omega t}\mathscr{\sigma}_{ab}^{\left(1\right)0}(t,t)\\
 & = & \frac{\left(-\boldsymbol{e}^{2}\right)}{2\pi i\omega_{1}}\int_{-\infty}^{+\infty}dte^{i\left(\omega-\omega_{1}\right)t}\int d\widetilde{\boldsymbol{k}}<|v^{ab}|>_{0}\\
 & = & \frac{\left(-\boldsymbol{e}^{2}\right)}{i\omega}\int d\widetilde{\boldsymbol{k}}<|v^{ab}|>_{0}\\
 & = & \frac{\left(-\boldsymbol{e}^{2}\right)}{\omega}\iint d\widetilde{\boldsymbol{k}}d\widetilde{E}f(E)Tr\left(v^{ab}g_{E}^{r-a}\right)\\
 & = & \left(-\frac{\boldsymbol{e}^{2}}{\hbar}\right)\frac{\hbar}{\omega}\iint d\widetilde{\boldsymbol{k}}d\widetilde{E}f(E)Tr\left(v^{ab}g_{E}^{r-a}\right)
\end{eqnarray*}

\begin{eqnarray*}
\mathscr{\sigma}_{ab}^{\left(1\right)1}(t,t^{\prime}) & = & -\frac{i}{\hbar}\int d\widetilde{\boldsymbol{k}}\Theta(t-t^{\prime})<|\left[R(t),\frac{\partial V_{1}^{I}(t^{\prime})}{\partial E(t^{\prime})}\right]|>_{0}\\
 & = & -\frac{\left(-\boldsymbol{e}^{2}\right)}{\hbar\omega_{1}}\int d\widetilde{\boldsymbol{k}}\Theta(t-t^{\prime})e^{-i\omega_{1}t^{\prime}}<|\left[v^{a}\left(t\right),v^{b}\left(t^{\prime}\right)\right]|>_{0}
\end{eqnarray*}

\begin{eqnarray*}
 &  & \mathscr{\sigma}_{ab}^{\left(1\right)1}(\omega=\omega_{1})\\
 & = & \frac{1}{2\pi}\iint_{-\infty}^{+\infty}dtdt^{\prime}e^{i\omega t}\mathscr{\sigma}_{ab}^{\left(1\right)1}(t,t^{\prime})\\
 & = & -\frac{\left(-\boldsymbol{e}^{2}\right)}{2\pi\hbar\omega_{1}}\int d\widetilde{\boldsymbol{k}}\iint_{-\infty}^{+\infty}dtdt^{\prime}e^{i(\omega t-\omega_{1}t^{\prime})}\Theta(t-t^{\prime})<|\left[v^{a}\left(t\right),v^{b}\left(t^{\prime}\right)\right]|>_{0}\\
 & = & -\frac{\left(-\boldsymbol{e}^{2}\right)}{2\pi\hbar\omega_{1}}\sum_{l,l^{\prime},m,m^{\prime}}\int d\widetilde{\boldsymbol{k}}\iint_{-\infty}^{+\infty}dtdt^{\prime}e^{i(\omega t-\omega_{1}t^{\prime})}v_{ll^{\prime}}^{a}v_{mm^{\prime}}^{b}\boldsymbol{\Theta(t-t^{\prime})<|\left[c_{l}^{\dagger}\left(t\right)c_{l^{\prime}}\left(t\right),c_{m}^{\dagger}\left(t^{\prime}\right)c_{m^{\prime}}\left(t^{\prime}\right)\right]|>_{0}}\\
 & = & -\frac{\left(-\boldsymbol{e}^{2}\right)}{2\pi\hbar\omega_{1}}\sum_{l,l^{\prime},m,m^{\prime}}\int d\widetilde{\boldsymbol{k}}\iint_{-\infty}^{+\infty}d(t-t^{\prime})dt^{\prime}e^{i\omega(t-t^{\prime})}e^{i(\omega-\omega_{1})t^{\prime}}\times\\
 &  & \,\,\,\,\,\,\,\,\,\,\,\,\,\,\,\,\,\,\,\,\,\,\,\,\,\,v_{ll^{\prime}}^{a}v_{mm^{\prime}}^{b}\left[\boldsymbol{g_{l^{\prime}m}^{r}(t-t^{\prime})g_{m^{\prime}l}^{<}(t^{\prime}-t)+g_{l^{\prime}m}^{<}(t-t^{\prime})g_{m^{\prime}l}^{a}(t^{\prime}-t)}\right]\\
 & = & -\frac{\left(-\boldsymbol{e}^{2}\right)}{\left(2\pi\right)^{3}\hbar\omega_{1}}\sum_{l,l^{\prime},m,m^{\prime}}\int d\widetilde{\boldsymbol{k}}\iint_{-\infty}^{+\infty}d(t-t^{\prime})dt^{\prime}\iint_{-\infty}^{+\infty}dE_{1}dE_{2}e^{-i(E_{1}-E_{2})(t-t^{\prime})/\hbar}e^{i\omega(t-t^{\prime})}e^{i(\omega-\omega_{1})t^{\prime}}\times\\
 &  & \,\,\,\,\,\,\,\,\,\,\,\,\,\,\,\,\,\,\,\,\,\,\,\,\,\,v_{ll^{\prime}}^{a}v_{mm^{\prime}}^{b}\left[g_{l^{\prime}m}^{r}(E_{1})g_{m^{\prime}l}^{<}(E_{2})+g_{l^{\prime}m}^{<}(E_{1})g_{m^{\prime}l}^{a}(E_{2})\right]\\
 & = & -\frac{\left(-\boldsymbol{e}^{2}\right)}{2\pi\omega}\sum_{l,l^{\prime},m,m^{\prime}}\iint d\widetilde{\boldsymbol{k}}dE_{2}v_{ll^{\prime}}^{a}v_{mm^{\prime}}^{b}\left[g_{l^{\prime}m}^{r}(E_{2}+\hbar\omega)g_{m^{\prime}l}^{<}(E_{2})+g_{l^{\prime}m}^{<}(E_{2}+\hbar\omega)g_{m^{\prime}l}^{a}(E_{2})\right]\\
 & = & \frac{\left(-\boldsymbol{e}^{2}\right)}{2\pi\omega}\iint d\widetilde{\boldsymbol{k}}dEf(E)Tr\left(v^{a}g_{E+\hbar\omega}^{r}v^{b}g_{E}^{r-a}+g_{E}^{r-a}v^{b}g_{E-\hbar\omega}^{a}v^{a}\right)\\
 & = & \left(-\frac{\boldsymbol{e}^{2}}{\hbar}\right)\frac{\hbar}{\omega}\iint d\widetilde{\boldsymbol{k}}d\widetilde{E}f(E)Tr\left(v^{a}g_{E+\hbar\omega}^{r}v^{b}g_{E}^{r-a}+g_{E}^{r-a}v^{b}g_{E-\hbar\omega}^{a}v^{a}\right)
\end{eqnarray*}

\begin{eqnarray*}
\mathscr{\sigma}_{abc}^{\left(2\right)11}(t,t^{\prime},t^{\prime\prime}) & = & (-\frac{i}{\hbar})^{2}\int d\widetilde{\boldsymbol{k}}\Theta(t-t^{\prime})\Theta(t^{\prime}-t^{\prime\prime})<|\left[\hat{R}(t),\left[\frac{\partial V_{1}^{I}(t^{\prime})}{\partial E(t^{\prime})},\frac{\partial V_{1}^{I}(t^{\prime\prime})}{\partial E(t^{\prime\prime})}\right]\right]|>_{0}\\
 & = & \frac{\left(-\boldsymbol{e}^{3}\right)e^{-i\omega_{1}t^{\prime}-i\omega_{2}t^{\prime\prime}}}{\hbar^{2}\omega_{1}\omega_{2}}\int d\widetilde{\boldsymbol{k}}\Theta(t-t^{\prime})\Theta(t^{\prime}-t^{\prime\prime})\int d\widetilde{\boldsymbol{k}}<|\left[\left[v^{a}\left(t\right),v^{b}\left(t^{\prime}\right)\right],v^{c}\left(t^{\prime\prime}\right)\right]|>_{0}
\end{eqnarray*}

\begin{eqnarray*}
 &  & \mathscr{\sigma}_{abc}^{\left(2\right)11}(\omega=\omega_{1}+\omega_{2})\\
 & = & \frac{1}{2\pi}\iiint_{-\infty}^{+\infty}dtdt^{\prime}dt^{\prime\prime}e^{i\omega t}\mathscr{\sigma}_{abc}^{\left(2\right)11}(t,t^{\prime},t^{\prime\prime})\\
 & = & \frac{\left(-\boldsymbol{e}^{3}\right)}{2\pi\hbar^{2}\omega_{1}\omega_{2}}\int d\widetilde{\boldsymbol{k}}\iiint_{-\infty}^{+\infty}dtdt^{\prime}dt^{\prime\prime}e^{i\omega t-i\omega_{1}t^{\prime}-i\omega_{2}t^{\prime\prime}}\Theta(t-t^{\prime})\Theta(t^{\prime}-t^{\prime\prime})<|\left[\left[v^{a}\left(t\right),v^{b}\left(t^{\prime}\right)\right],v^{c}\left(t^{\prime\prime}\right)\right]|>_{0}\\
 & = & \frac{\left(-\boldsymbol{e}^{3}\right)}{2\pi\hbar^{2}\omega_{1}\omega_{2}}\sum_{l,l^{\prime},m,m^{\prime},n,n^{\prime}}\int d\widetilde{\boldsymbol{k}}\iiint_{-\infty}^{+\infty}dtdt^{\prime}dt^{\prime\prime}e^{i\omega t-i\omega_{1}t^{\prime}-i\omega_{2}t^{\prime\prime}}v_{ll^{\prime}}^{a}v_{mm^{\prime}}^{b}v_{nn^{\prime}}^{c}\times\\
 &  & \,\,\,\,\,\,\,\,\,\,\,\,\,\,\,\,\,\,\,\,\,\,\,\,\,\,\,\,\,\,\,\,\,\,\boldsymbol{\Theta(t-t^{\prime})\Theta(t^{\prime}-t^{\prime\prime})<|\left[\left[c_{l}^{\dagger}\left(t\right)c_{l^{\prime}}\left(t\right),c_{m}^{\dagger}\left(t^{\prime}\right)c_{m^{\prime}}\left(t^{\prime}\right)\right],c_{n}^{\dagger}\left(t^{\prime\prime}\right)c_{n^{\prime}}\left(t^{\prime\prime}\right)\right]|>_{0}}\\
 & = & +\frac{i\left(-\boldsymbol{e}^{3}\right)}{2\pi\hbar^{2}\omega_{1}\omega_{2}}\sum_{l,l^{\prime},m,m^{\prime},n,n^{\prime}}\int d\widetilde{\boldsymbol{k}}\iiint_{-\infty}^{+\infty}dtdt^{\prime}dt^{\prime\prime}e^{i\omega t-i\omega_{1}t^{\prime}-i\omega_{2}t^{\prime\prime}}v_{ll^{\prime}}^{a}v_{mm^{\prime}}^{b}v_{nn^{\prime}}^{c}\times\\
1 &  & \,\,\,\,\,\,\,\,\,\,\,\,\,\,\,\,\,\,\,\,\,\,\,\,\,\,\,\,\,\,\,\,\,\,\,\,\,\,\,\,\,\,\,\,\,\,\,\,\,\,\boldsymbol{\left\{ g_{l^{\prime}m}^{r}(t-t^{\prime})g_{m^{\prime}n}^{r}(t^{\prime}-t^{\prime\prime})g_{n^{\prime}l}^{<}(t^{\prime\prime}-t)\right.}\\
2 &  & \,\,\,\,\,\,\,\,\,\,\,\,\,\,\,\,\,\,\,\,\,\,\,\,\,\,\,\,\,\,\,\,\,\,\,\,\,\,\,\,\,\,\,\,\,\,\,\,\,\,+\boldsymbol{\Theta(t^{\prime}-t^{\prime\prime})g_{l^{\prime}m}^{r}(t-t^{\prime})g_{m^{\prime}n}^{<}(t^{\prime}-t^{\prime\prime})\left[g_{n^{\prime}l}^{<}(t^{\prime\prime}-t)-g_{n^{\prime}l}^{>}(t^{\prime\prime}-t)\right]}\\
3 &  & \,\,\,\,\,\,\,\,\,\,\,\,\,\,\,\,\,\,\,\,\,\,\,\,\,\,\,\,\,\,\,\,\,\,\,\,\,\,\,\,\,\,\,\,\,\,\,\,\,\,\boldsymbol{+g_{m^{\prime}l}^{a}(t^{\prime}-t)g_{n^{\prime}m}^{a}(t^{\prime\prime}-t^{\prime})g_{l^{\prime}n}^{<}(t-t^{\prime\prime})}\\
4 &  & \,\,\,\,\,\,\,\,\,\,\,\,\,\,\,\,\,\,\,\,\,\,\,\,\,\,\,\,\,\,\,\,\,\,\,\,\,\,\,\,\,\,\,\,\,\,\,\,\,\,\boldsymbol{\left.\Theta(t^{\prime}-t^{\prime\prime})g_{m^{\prime}l}^{a}(t^{\prime}-t)g_{n^{\prime}m}^{<}(t^{\prime\prime}-t^{\prime})\left[g_{l^{\prime}n}^{>}(t-t^{\prime\prime})-g_{l^{\prime}n}^{<}(t-t^{\prime\prime})\right]\right\} }\\
\\ &  & \mathscr{\sigma}_{abc1}^{\left(2\right)11}(\omega=\omega_{1}+\omega_{2})\\
 & = & +\frac{i\left(-\boldsymbol{e}^{3}\right)}{2\pi\hbar^{2}\omega_{1}\omega_{2}}\sum_{l,l^{\prime},m,m^{\prime},n,n^{\prime}}\int d\widetilde{\boldsymbol{k}}\iiint_{-\infty}^{+\infty}dtdt^{\prime}dt^{\prime\prime}\times\\
 &  & \,\,\,\,\,\,\,\,\,\,\,\,\,\,\,\,\,\,\,\,\,\,\,e^{i\omega t-i\omega_{1}t^{\prime}-i\omega_{2}t^{\prime\prime}}v_{ll^{\prime}}^{a}v_{mm^{\prime}}^{b}v_{nn^{\prime}}^{c}\left[g_{l^{\prime}m}^{r}(t-t^{\prime})g_{m^{\prime}n}^{r}(t^{\prime}-t^{\prime\prime})g_{n^{\prime}l}^{<}(t^{\prime\prime}-t)\right]\\
 & = & -\frac{i\left(-\boldsymbol{e}^{3}\right)}{2\pi\omega_{1}\omega_{2}}\int d\widetilde{\boldsymbol{k}}\int dEf(E)Tr\left(v^{a}g_{E+\hbar\omega}^{r}v^{b}g_{E+\hbar\omega_{2}}^{r}v^{c}g_{E}^{r-a}\right)\\
\\ &  & \mathscr{\sigma}_{abc2}^{\left(2\right)11}(\omega=\omega_{1}+\omega_{2})\\
 & = & +\frac{i\left(-\boldsymbol{e}^{3}\right)}{2\pi\hbar^{2}\omega_{1}\omega_{2}}\sum_{l,l^{\prime},m,m^{\prime},n,n^{\prime}}\int d\widetilde{\boldsymbol{k}}\iiint_{-\infty}^{+\infty}dtdt^{\prime}dt^{\prime\prime}e^{i\omega t-i\omega_{1}t^{\prime}-i\omega_{2}t^{\prime\prime}}\times\\
 &  & \,\,\,\,\,\,\,\,\,\,\,\,\,\,\,\,\,\,\,\,\,\,\,v_{ll^{\prime}}^{a}v_{mm^{\prime}}^{b}v_{nn^{\prime}}^{c}\Theta(t^{\prime}-t^{\prime\prime})\left\{ g_{l^{\prime}m}^{r}(t-t^{\prime})g_{m^{\prime}n}^{<}(t^{\prime}-t^{\prime\prime})\left[g_{n^{\prime}l}^{<}(t^{\prime\prime}-t)-g_{n^{\prime}l}^{>}(t^{\prime\prime}-t)\right]\right\} \\
 & = & -\frac{\left(-\boldsymbol{e}^{3}\right)}{(2\pi)^{2}\omega_{1}\omega_{2}}\int d\widetilde{\boldsymbol{k}}\iint dE^{\prime}dEf(E^{\prime})Tr\left(\frac{v^{a}g_{E+\hbar\omega}^{r}v^{b}g_{E^{\prime}}^{r-a}v^{c}g_{E}^{r-a}}{E+\hbar\omega_{2}-E^{\prime}+i\eta}\right)\\
\\ &  & \mathscr{\sigma}_{abc3}^{\left(2\right)11}(\omega=\omega_{1}+\omega_{2})\\
 & = & +\frac{i\left(-\boldsymbol{e}^{3}\right)}{2\pi\hbar^{2}\omega_{1}\omega_{2}}\sum_{l,l^{\prime},m,m^{\prime},n,n^{\prime}}\int d\widetilde{\boldsymbol{k}}\iiint_{-\infty}^{+\infty}dtdt^{\prime}dt^{\prime\prime}e^{i\omega t-i\omega_{1}t^{\prime}-i\omega_{2}t^{\prime\prime}}\times\\
 &  & \,\,\,\,\,\,\,\,\,\,\,\,\,\,\,\,\,\,\,\,\,\,\,v_{ll^{\prime}}^{a}v_{mm^{\prime}}^{b}v_{nn^{\prime}}^{c}\left[g_{m^{\prime}l}^{a}(t^{\prime}-t)g_{n^{\prime}m}^{a}(t^{\prime\prime}-t^{\prime})g_{l^{\prime}n}^{<}(t-t^{\prime\prime})\right]\\
 & = & -\frac{i\left(-\boldsymbol{e}^{3}\right)}{2\pi\omega_{1}\omega_{2}}\int d\widetilde{\boldsymbol{k}}\int dEf(E)Tr\left(g_{E}^{r-a}v^{c}g_{E-\hbar\omega_{2}}^{a}v^{b}g_{E-\hbar\omega}^{a}v^{a}\right)\\
\\ &  & \mathscr{\sigma}_{abc4}^{\left(2\right)11}(\omega=\omega_{1}+\omega_{2})\\
 & = & +\frac{i\left(-\boldsymbol{e}^{3}\right)}{2\pi\hbar^{2}\omega_{1}\omega_{2}}\sum_{l,l^{\prime},m,m^{\prime},n,n^{\prime}}\int d\widetilde{\boldsymbol{k}}\iiint_{-\infty}^{+\infty}dtdt^{\prime}dt^{\prime\prime}e^{i\omega t-i\omega_{1}t^{\prime}-i\omega_{2}t^{\prime\prime}}\times\\
 &  & \,\,\,\,\,\,\,\,\,\,\,v_{ll^{\prime}}^{a}v_{mm^{\prime}}^{b}v_{nn^{\prime}}^{c}\Theta(t^{\prime}-t^{\prime\prime})\left\{ g_{m^{\prime}l}^{a}(t^{\prime}-t)g_{n^{\prime}m}^{<}(t^{\prime\prime}-t^{\prime})\left[g_{l^{\prime}n}^{>}(t-t^{\prime\prime})-g_{l^{\prime}n}^{<}(t-t^{\prime\prime})\right]\right\} \\
 & = & -\frac{\left(-\boldsymbol{e}^{3}\right)}{(2\pi)^{2}\omega_{1}\omega_{2}}\int d\widetilde{\boldsymbol{k}}\iint dEdE^{\prime}f(E^{\prime})Tr\left(\frac{g_{E}^{r-a}v^{c}g_{E^{\prime}}^{r-a}v^{b}g_{E-\hbar\omega}^{a}v^{a}}{E-\hbar\omega_{2}-E^{\prime}-i\eta}\right)\\
\\\\ &  & \mathscr{\sigma}_{abc}^{\left(2\right)11}(\omega=\omega_{1}+\omega_{2})\\
 & = & \mathscr{\sigma}_{abc1}^{\left(2\right)11}(\omega=\omega_{1}+\omega_{2})+\mathscr{\sigma}_{abc2}^{\left(2\right)11}(\omega=\omega_{1}+\omega_{2})+\mathscr{\sigma}_{abc3}^{\left(2\right)11}(\omega=\omega_{1}+\omega_{2})+\mathscr{\sigma}_{abc4}^{\left(2\right)11}(\omega=\omega_{1}+\omega_{2})\\
 & = & \left(-\frac{\boldsymbol{e}^{3}}{\hbar^{2}}\right)\frac{\hbar^{2}}{\omega_{1}\omega_{2}}\iiint d\widetilde{\boldsymbol{k}}d\widetilde{E}d\widetilde{E}^{\prime}\left[f(E)-f(E^{\prime})\right]Tr\left(\frac{v^{a}g_{E+\hbar\omega}^{r}v^{b}g_{E^{\prime}}^{r-a}v^{c}g_{E}^{r-a}}{E+\hbar\omega_{2}-E^{\prime}+i\eta}+\frac{g_{E}^{r-a}v^{c}g_{E^{\prime}}^{r-a}v^{b}g_{E-\hbar\omega}^{a}v^{a}}{E-\hbar\omega_{2}-E^{\prime}-i\eta}\right)
\end{eqnarray*}

\begin{eqnarray*}
\mathscr{\sigma}_{abcd}^{\left(3\right)111}(t,t^{\prime},t^{\prime\prime},t^{\prime\prime\prime}) & = & (-\frac{i}{\hbar})^{3}\Theta(t-t^{\prime})\Theta(t^{\prime}-t^{\prime\prime})\Theta(t^{\prime\prime}-t^{\prime\prime\prime})<|\left[\hat{R}(t),\left[\frac{\partial V_{1}^{I}(t^{\prime})}{\partial E(t^{\prime})},\left[\frac{\partial V_{1}^{I}(t^{\prime\prime})}{\partial E(t^{\prime\prime})},\frac{\partial V_{1}^{I}(t^{\prime\prime\prime})}{\partial E(t^{\prime\prime\prime})}\right]\right]\right]|>_{0}\\
 & = & -\frac{\left(-\boldsymbol{e}^{4}\right)e^{-i\omega_{1}t^{\prime}}e^{-i\omega_{2}t^{\prime\prime}}e^{-i\omega_{3}t^{\prime\prime\prime}}}{\hbar^{3}\omega_{1}\omega_{2}\omega_{3}}\int d\widetilde{\boldsymbol{k}}\Theta(t-t^{\prime})\Theta(t^{\prime}-t^{\prime\prime})\Theta(t^{\prime\prime}-t^{\prime\prime\prime})\times\\
 &  & \,\,\,\,\,\,\,\,\,\,\,\,\,\,\,<|\left[\left[\left[v^{a}(t),v^{b}(t^{\prime})\right],v^{c}(t^{\prime\prime})\right],v^{d}(t^{\prime\prime\prime})\right]|>_{0}\\
\end{eqnarray*}

\begin{eqnarray*}
 &  & \mathscr{\sigma}_{abcd}^{\left(3\right)111}(\omega=\omega_{1}+\omega_{2}+\omega_{3})\\
 & = & -\frac{\left(-\boldsymbol{e}^{4}\right)}{2\pi\hbar^{3}\omega_{1}\omega_{2}\omega_{3}}\int d\widetilde{\boldsymbol{k}}\iiiint dtdt^{\prime}dt^{\prime\prime}dt^{\prime\prime\prime}\Theta(t-t^{\prime})\Theta(t^{\prime}-t^{\prime\prime})\Theta(t^{\prime\prime}-t^{\prime\prime\prime})\times\\
 &  & \,\,\,\,\,\,\,\,\,\,\,e^{i\omega t-i\omega_{1}t^{\prime}}e^{-i\omega_{2}t^{\prime\prime}}e^{-i\omega_{3}t^{\prime\prime\prime}}<|\left[\left[\left[v^{a}(t),v^{b}(t^{\prime})\right],v^{c}(t^{\prime\prime})\right],v^{d}(t^{\prime\prime\prime})\right]|>_{0}\\
 & = & -\frac{\left(-\boldsymbol{e}^{4}\right)}{2\pi\hbar^{3}\omega_{1}\omega_{2}\omega_{3}}\sum_{l,l^{\prime},m,m^{\prime},n,n^{\prime},p,p^{\prime}}\int d\widetilde{\boldsymbol{k}}\iiiint dtdt^{\prime}dt^{\prime\prime}dt^{\prime\prime\prime}e^{i\omega t-i\omega_{1}t^{\prime}}e^{-i\omega_{2}t^{\prime\prime}}e^{-i\omega_{3}t^{\prime\prime\prime}}v_{ll^{\prime}}^{a}v_{mm^{\prime}}^{b}v_{nn^{\prime}}^{c}v_{pp^{\prime}}^{d}\times\\
 &  & \boldsymbol{\Theta(t-t^{\prime})\Theta(t^{\prime}-t^{\prime\prime})\Theta(t^{\prime\prime}-t^{\prime\prime\prime})<|\left[\left[\left[c_{l}^{\dagger}\left(t\right)c_{l^{\prime}}\left(t\right),c_{m}^{\dagger}\left(t^{\prime}\right)c_{m^{\prime}}\left(t^{\prime}\right)\right],c_{n}^{\dagger}\left(t^{\prime\prime}\right)c_{n^{\prime}}\left(t^{\prime\prime}\right)\right],c_{p}^{\dagger}\left(t^{\prime\prime\prime}\right)c_{p^{\prime}}\left(t^{\prime\prime\prime}\right)\right]|>_{0}}\\
 & = & -\frac{\left(-\boldsymbol{e}^{4}\right)}{2\pi\hbar^{3}\omega_{1}\omega_{2}\omega_{3}}\sum_{l,l^{\prime},m,m^{\prime},n,n^{\prime},p,p^{\prime}}\int d\widetilde{\boldsymbol{k}}\iiiint dtdt^{\prime}dt^{\prime\prime}dt^{\prime\prime\prime}e^{i\omega t-i\omega_{1}t^{\prime}}e^{-i\omega_{2}t^{\prime\prime}}e^{-i\omega_{3}t^{\prime\prime\prime}}v_{ll^{\prime}}^{a}v_{mm^{\prime}}^{b}v_{nn^{\prime}}^{c}v_{pp^{\prime}}^{d}\times\\
1 &  & \boldsymbol{\left\{ g_{l^{\prime}m}^{r}(t-t^{\prime})g_{m^{\prime}n}^{r}(t^{\prime}-t^{\prime\prime})g_{n^{\prime}p}^{r}(t^{\prime\prime}-t^{\prime\prime\prime})g_{p^{\prime}l}^{<}(t^{\prime\prime\prime}-t)\right.}\\
2 &  & \boldsymbol{+\Theta(t^{\prime\prime}-t^{\prime\prime\prime})g_{l^{\prime}m}^{r}(t-t^{\prime})g_{m^{\prime}n}^{r}(t^{\prime}-t^{\prime\prime})g_{n^{\prime}p}^{<}(t^{\prime\prime}-t^{\prime\prime\prime})\left[g_{p^{\prime}l}^{<}(t^{\prime\prime\prime}-t)-g_{p^{\prime}l}^{>}(t^{\prime\prime\prime}-t)\right]}\\
3 &  & \boldsymbol{+\Theta(t^{\prime}-t^{\prime\prime})g_{l^{\prime}m}^{r}(t-t^{\prime})g_{m^{\prime}p}^{<}(t^{\prime}-t^{\prime\prime\prime})g_{p^{\prime}n}^{a}(t^{\prime\prime\prime}-t^{\prime\prime})\left[g_{n^{\prime}l}^{<}(t^{\prime\prime}-t)-g_{n^{\prime}l}^{>}(t^{\prime\prime}-t)\right]}\\
4 &  & \boldsymbol{+\Theta(t^{\prime}-t^{\prime\prime})\Theta(t^{\prime\prime}-t^{\prime\prime\prime})g_{l^{\prime}m}^{r}(t-t^{\prime})\left[g_{m^{\prime}p}^{>}(t^{\prime}-t^{\prime\prime\prime})-g_{m^{\prime}p}^{<}(t^{\prime}-t^{\prime\prime\prime})\right]}\\
 &  & \,\,\,\,\,\,\,\,\,\,\,\,\,\,\,\,\,\,\,\,\,\,\,\,\,\,\,\,\,\,\,\,\,\,\,\,\,\,\,\,\,\,\,\,\,\,\boldsymbol{\times g_{p^{\prime}n}^{<}(t^{\prime\prime\prime}-t^{\prime\prime})\left[g_{n^{\prime}l}^{<}(t^{\prime\prime}-t)-g_{n^{\prime}l}^{>}(t^{\prime\prime}-t)\right]}\\
5 &  & \boldsymbol{+\Theta(t^{\prime}-t^{\prime\prime})\Theta(t^{\prime\prime}-t^{\prime\prime\prime})g_{m^{\prime}l}^{a}(t^{\prime}-t)\left[g_{p^{\prime}m}^{<}(t^{\prime\prime\prime}-t^{\prime})-g_{p^{\prime}m}^{>}(t^{\prime\prime\prime}-t^{\prime})\right]}\\
 &  & \,\,\,\,\,\,\,\,\,\,\,\,\,\,\,\,\,\,\,\,\,\,\,\,\,\,\,\,\,\,\,\,\,\,\,\,\,\,\,\,\,\,\,\,\,\,\boldsymbol{\times g_{n^{\prime}p}^{<}(t^{\prime\prime}-t^{\prime\prime\prime})\left[g_{l^{\prime}n}^{>}(t-t^{\prime\prime})-g_{l^{\prime}n}^{<}(t-t^{\prime\prime})\right]}\\
6 &  & \boldsymbol{+\Theta(t^{\prime}-t^{\prime\prime})g_{m^{\prime}l}^{a}(t^{\prime}-t)g_{p^{\prime}m}^{<}(t^{\prime\prime\prime}-t^{\prime})g_{n^{\prime}p}^{r}(t^{\prime\prime}-t^{\prime\prime\prime})\left[g_{l^{\prime}n}^{>}(t-t^{\prime\prime})-g_{l^{\prime}n}^{<}(t-t^{\prime\prime})\right]}\\
7 &  & \boldsymbol{+\Theta(t^{\prime\prime}-t^{\prime\prime\prime})g_{m^{\prime}l}^{a}(t^{\prime}-t)g_{n^{\prime}m}^{a}(t^{\prime\prime}-t^{\prime})g_{p^{\prime}n}^{<}(t^{\prime\prime\prime}-t^{\prime\prime})\left[g_{l^{\prime}p}^{>}(t-t^{\prime\prime\prime})-g_{l^{\prime}p}^{<}(t-t^{\prime\prime\prime})\right]}\\
8 &  & \boldsymbol{\left.+g_{m^{\prime}l}^{a}(t^{\prime}-t)g_{n^{\prime}m}^{a}(t^{\prime\prime}-t^{\prime})g_{p^{\prime}n}^{a}(t^{\prime\prime\prime}-t^{\prime\prime})g_{l^{\prime}p}^{<}(t-t^{\prime\prime\prime})\right\} }\\
\\ &  & \mathscr{\sigma}_{abcd1}^{\left(3\right)111}(\omega=\omega_{1}+\omega_{2}+\omega_{3})\\
 & = & -\frac{\left(-\boldsymbol{e}^{4}\right)}{2\pi\hbar^{3}\omega_{1}\omega_{2}\omega_{3}}\sum_{l,l^{\prime},m,m^{\prime},n,n^{\prime},p,p^{\prime}}\int d\widetilde{\boldsymbol{k}}\iiiint dtdt^{\prime}dt^{\prime\prime}dt^{\prime\prime\prime}e^{i\omega t-i\omega_{1}t^{\prime}}e^{-i\omega_{2}t^{\prime\prime}}e^{-i\omega_{3}t^{\prime\prime\prime}}\\
 &  & \,\,\,\,\,\,\,\,\,\,\,\,\,\times v_{ll^{\prime}}^{a}v_{mm^{\prime}}^{b}v_{nn^{\prime}}^{c}v_{pp^{\prime}}^{d}g_{l^{\prime}m}^{r}(t-t^{\prime})g_{m^{\prime}n}^{r}(t^{\prime}-t^{\prime\prime})g_{n^{\prime}p}^{r}(t^{\prime\prime}-t^{\prime\prime\prime})g_{p^{\prime}l}^{<}(t^{\prime\prime\prime}-t)\\
 & = & \frac{\left(-\boldsymbol{e}^{4}\right)}{2\pi\omega_{1}\omega_{2}\omega_{3}}\iint d\widetilde{\boldsymbol{k}}dEf(E)Tr\left(v^{a}g_{E+\hbar\omega}^{r}v^{b}g_{E+\hbar\omega_{2}+\hbar\omega_{3}}^{r}v^{c}g_{E+\hbar\omega_{3}}^{r}v^{d}g_{E}^{r-a}\right)\\
\\ &  & \mathscr{\sigma}_{abcd2}^{\left(3\right)111}(\omega=\omega_{1}+\omega_{2}+\omega_{3})\\
 & = & -\frac{\left(-\boldsymbol{e}^{4}\right)}{2\pi\hbar^{3}\omega_{1}\omega_{2}\omega_{3}}\sum_{l,l^{\prime},m,m^{\prime},n,n^{\prime},p,p^{\prime}}\int d\widetilde{\boldsymbol{k}}\iiiint dtdt^{\prime}dt^{\prime\prime}dt^{\prime\prime\prime}e^{i\omega t-i\omega_{1}t^{\prime}}e^{-i\omega_{2}t^{\prime\prime}}e^{-i\omega_{3}t^{\prime\prime\prime}}v_{ll^{\prime}}^{a}v_{mm^{\prime}}^{b}v_{nn^{\prime}}^{c}v_{pp^{\prime}}^{d}\\
 &  & \,\,\,\,\,\,\,\,\,\,\,\,\,\times\Theta(t^{\prime\prime}-t^{\prime\prime\prime})g_{l^{\prime}m}^{r}(t-t^{\prime})g_{m^{\prime}n}^{r}(t^{\prime}-t^{\prime\prime})g_{n^{\prime}p}^{<}(t^{\prime\prime}-t^{\prime\prime\prime})\left[g_{p^{\prime}l}^{<}(t^{\prime\prime\prime}-t)-g_{p^{\prime}l}^{>}(t^{\prime\prime\prime}-t)\right]\\
 & = & -\frac{i\left(-\boldsymbol{e}^{4}\right)}{(2\pi)^{2}\omega_{1}\omega_{2}\omega_{3}}\iiint d\widetilde{\boldsymbol{k}}dEdE^{\prime}f(E^{\prime})Tr\left(\frac{v^{a}g_{E+\hbar\omega}^{r}v^{b}g_{E+\hbar\omega_{2}+\hbar\omega_{3}}^{r}v^{c}g_{E^{\prime}}^{r-a}v^{d}g_{E}^{r-a}}{E+\hbar\omega_{3}-E^{\prime}+i\eta}\right)\\
\\ &  & \mathscr{\sigma}_{abcd3}^{\left(3\right)111}(\omega=\omega_{1}+\omega_{2}+\omega_{3})\\
 & = & -\frac{\left(-\boldsymbol{e}^{4}\right)}{2\pi\hbar^{3}\omega_{1}\omega_{2}\omega_{3}}\sum_{l,l^{\prime},m,m^{\prime},n,n^{\prime},p,p^{\prime}}\int d\widetilde{\boldsymbol{k}}\iiiint dtdt^{\prime}dt^{\prime\prime}dt^{\prime\prime\prime}e^{i\omega t-i\omega_{1}t^{\prime}}e^{-i\omega_{2}t^{\prime\prime}}e^{-i\omega_{3}t^{\prime\prime\prime}}v_{ll^{\prime}}^{a}v_{mm^{\prime}}^{b}v_{nn^{\prime}}^{c}v_{pp^{\prime}}^{d}\\
 &  & \,\,\,\,\,\,\,\,\,\,\,\,\,\times\Theta(t^{\prime}-t^{\prime\prime})g_{l^{\prime}m}^{r}(t-t^{\prime})g_{m^{\prime}p}^{<}(t^{\prime}-t^{\prime\prime\prime})g_{p^{\prime}n}^{a}(t^{\prime\prime\prime}-t^{\prime\prime})\left[g_{n^{\prime}l}^{<}(t^{\prime\prime}-t)-g_{n^{\prime}l}^{>}(t^{\prime\prime}-t)\right]\\
 & = & -\frac{\left(-\boldsymbol{e}^{4}\right)}{(2\pi)^{2}\omega_{1}\omega_{2}\omega_{3}}\iiint d\widetilde{\boldsymbol{k}}dEdE^{\prime}f(E^{\prime})Tr\left(\frac{v^{a}g_{E+\hbar\omega}^{r}v^{b}g_{E^{\prime}}^{r-a}v^{d}g_{E^{\prime}-\hbar\omega_{3}}^{a}v^{c}g_{E}^{r-a}}{E-E^{\prime}+\hbar\omega_{2}+\hbar\omega_{3}+i\eta}\right)\\
\end{eqnarray*}

\begin{eqnarray*}
 &  & \mathscr{\sigma}_{abcd4}^{\left(3\right)111}(\omega=\omega_{1}+\omega_{2}+\omega_{3})\\
 & = & -\frac{\left(-\boldsymbol{e}^{4}\right)}{2\pi\hbar^{3}\omega_{1}\omega_{2}\omega_{3}}\sum_{l,l^{\prime},m,m^{\prime},n,n^{\prime},p,p^{\prime}}\int d\widetilde{\boldsymbol{k}}\iiiint dtdt^{\prime}dt^{\prime\prime}dt^{\prime\prime\prime}e^{i\omega t-i\omega_{1}t^{\prime}}e^{-i\omega_{2}t^{\prime\prime}}e^{-i\omega_{3}t^{\prime\prime\prime}}v_{ll^{\prime}}^{a}v_{mm^{\prime}}^{b}v_{nn^{\prime}}^{c}v_{pp^{\prime}}^{d}\\
 &  & \,\,\,\,\,\,\,\,\,\,\,\,\,\times\Theta(t^{\prime}-t^{\prime\prime})\Theta(t^{\prime\prime}-t^{\prime\prime\prime})g_{l^{\prime}m}^{r}(t-t^{\prime})\left[g_{m^{\prime}p}^{>}(t^{\prime}-t^{\prime\prime\prime})-g_{m^{\prime}p}^{<}(t^{\prime}-t^{\prime\prime\prime})\right]g_{p^{\prime}n}^{<}(t^{\prime\prime\prime}-t^{\prime\prime})\left[g_{n^{\prime}l}^{<}(t^{\prime\prime}-t)-g_{n^{\prime}l}^{>}(t^{\prime\prime}-t)\right]\\
 & = & -\frac{\left(-\boldsymbol{e}^{4}\right)}{(2\pi)^{3}\omega_{1}\omega_{2}\omega_{3}}\iiiint d\widetilde{\boldsymbol{k}}dEdE^{\prime}dE^{\prime\prime}f(E^{\prime\prime})Tr\left[\frac{v^{a}g_{E+\hbar\omega}^{r}v^{b}g_{E^{\prime}}^{r-a}v^{d}g_{E^{\prime\prime}}^{r-a}v^{c}g_{E}^{r-a}}{\left(E^{\prime}-\hbar\omega_{3}-E^{\prime\prime}-i\eta\right)\left(E+\hbar\omega_{2}+\hbar\omega_{3}-E^{\prime}+i\eta\right)}\right]\\
\end{eqnarray*}

\begin{eqnarray*}
 &  & \mathscr{\sigma}_{abcd5}^{\left(3\right)111}(\omega=\omega_{1}+\omega_{2}+\omega_{3})\\
 & = & -\frac{\left(-\boldsymbol{e}^{4}\right)}{2\pi\hbar^{3}\omega_{1}\omega_{2}\omega_{3}}\sum_{l,l^{\prime},m,m^{\prime},n,n^{\prime},p,p^{\prime}}\int d\widetilde{\boldsymbol{k}}\iiiint dtdt^{\prime}dt^{\prime\prime}dt^{\prime\prime\prime}e^{i\omega t-i\omega_{1}t^{\prime}}e^{-i\omega_{2}t^{\prime\prime}}e^{-i\omega_{3}t^{\prime\prime\prime}}v_{ll^{\prime}}^{a}v_{mm^{\prime}}^{b}v_{nn^{\prime}}^{c}v_{pp^{\prime}}^{d}\\
 &  & \,\,\,\,\,\,\,\,\,\,\,\,\,\times\Theta(t^{\prime}-t^{\prime\prime})\Theta(t^{\prime\prime}-t^{\prime\prime\prime})g_{m^{\prime}l}^{a}(t^{\prime}-t)\left[g_{p^{\prime}m}^{<}(t^{\prime\prime\prime}-t^{\prime})-g_{p^{\prime}m}^{>}(t^{\prime\prime\prime}-t^{\prime})\right]g_{n^{\prime}p}^{<}(t^{\prime\prime}-t^{\prime\prime\prime})\left[g_{l^{\prime}n}^{>}(t-t^{\prime\prime})-g_{l^{\prime}n}^{<}(t-t^{\prime\prime})\right]\\
 & = & -\frac{\left(-\boldsymbol{e}^{4}\right)}{(2\pi)^{3}\omega_{1}\omega_{2}\omega_{3}}\iiiint d\widetilde{\boldsymbol{k}}dEdE^{\prime}dE^{\prime\prime}f(E^{\prime\prime})Tr\left[\frac{g_{E}^{r-a}v^{c}g_{E^{\prime\prime}}^{r-a}v^{d}g_{E^{\prime}}^{r-a}v^{b}g_{E-\hbar\omega}^{a}v^{a}}{\left(E^{\prime}+\hbar\omega_{3}-E^{\prime\prime}+i\eta\right)\left(E-\hbar\omega_{2}-\hbar\omega_{3}-E^{\prime}-i\eta\right)}\right]\\
\end{eqnarray*}

\begin{eqnarray*}
 &  & \mathscr{\sigma}_{abcd6}^{\left(3\right)111}(\omega=\omega_{1}+\omega_{2}+\omega_{3})\\
 & = & -\frac{\left(-\boldsymbol{e}^{4}\right)}{2\pi\hbar^{3}\omega_{1}\omega_{2}\omega_{3}}\sum_{l,l^{\prime},m,m^{\prime},n,n^{\prime},p,p^{\prime}}\int d\widetilde{\boldsymbol{k}}\iiiint dtdt^{\prime}dt^{\prime\prime}dt^{\prime\prime\prime}e^{i\omega t-i\omega_{1}t^{\prime}}e^{-i\omega_{2}t^{\prime\prime}}e^{-i\omega_{3}t^{\prime\prime\prime}}v_{ll^{\prime}}^{a}v_{mm^{\prime}}^{b}v_{nn^{\prime}}^{c}v_{pp^{\prime}}^{d}\\
 &  & \,\,\,\,\,\,\,\,\,\,\,\,\,\times\Theta(t^{\prime}-t^{\prime\prime})g_{m^{\prime}l}^{a}(t^{\prime}-t)g_{p^{\prime}m}^{<}(t^{\prime\prime\prime}-t^{\prime})g_{n^{\prime}p}^{r}(t^{\prime\prime}-t^{\prime\prime\prime})\left[g_{l^{\prime}n}^{>}(t-t^{\prime\prime})-g_{l^{\prime}n}^{<}(t-t^{\prime\prime})\right]\\
 & = & -\frac{i\left(-\boldsymbol{e}^{4}\right)}{(2\pi)^{2}\omega_{1}\omega_{2}\omega_{3}}\iiint d\widetilde{\boldsymbol{k}}dEdE^{\prime}f(E^{\prime})Tr\left(\frac{g_{E}^{r-a}v^{c}g_{E^{\prime}+\hbar\omega_{3}}^{r}v^{d}g_{E^{\prime}}^{r-a}v^{b}g_{E-\hbar\omega}^{a}v^{a}}{E-\hbar\omega_{2}-\hbar\omega_{3}-E^{\prime}-i\eta}\right)\\
\\ &  & \mathscr{\sigma}_{abcd7}^{\left(3\right)111}(\omega=\omega_{1}+\omega_{2}+\omega_{3})\\
 & = & -\frac{\left(-\boldsymbol{e}^{4}\right)}{2\pi\hbar^{3}\omega_{1}\omega_{2}\omega_{3}}\sum_{l,l^{\prime},m,m^{\prime},n,n^{\prime},p,p^{\prime}}\int d\widetilde{\boldsymbol{k}}\iiiint dtdt^{\prime}dt^{\prime\prime}dt^{\prime\prime\prime}e^{i\omega t-i\omega_{1}t^{\prime}}e^{-i\omega_{2}t^{\prime\prime}}e^{-i\omega_{3}t^{\prime\prime\prime}}v_{ll^{\prime}}^{a}v_{mm^{\prime}}^{b}v_{nn^{\prime}}^{c}v_{pp^{\prime}}^{d}\\
 &  & \,\,\,\,\,\,\,\,\,\,\,\,\,\times\Theta(t^{\prime\prime}-t^{\prime\prime\prime})g_{m^{\prime}l}^{a}(t^{\prime}-t)g_{n^{\prime}m}^{a}(t^{\prime\prime}-t^{\prime})g_{p^{\prime}n}^{<}(t^{\prime\prime\prime}-t^{\prime\prime})\left[g_{l^{\prime}p}^{>}(t-t^{\prime\prime\prime})-g_{l^{\prime}p}^{<}(t-t^{\prime\prime\prime})\right]\\
 & = & -\frac{i\left(-\boldsymbol{e}^{4}\right)}{(2\pi)^{2}\omega_{1}\omega_{2}\omega_{3}}\iiint d\widetilde{\boldsymbol{k}}dEdE^{\prime}f(E^{\prime})Tr\left(\frac{g_{E}^{r-a}v^{d}g_{E^{\prime}}^{r-a}v^{c}g_{E-\hbar\omega_{2}-\hbar\omega_{3}}^{a}v^{b}g_{E-\hbar\omega}^{a}v^{a}}{E-\hbar\omega_{3}-E^{\prime}-i\eta}\right)\\
\\ &  & \mathscr{\sigma}_{abcd8}^{\left(3\right)111}(\omega=\omega_{1}+\omega_{2}+\omega_{3})\\
 & = & -\frac{\left(-\boldsymbol{e}^{4}\right)}{2\pi\hbar^{3}\omega_{1}\omega_{2}\omega_{3}}\sum_{l,l^{\prime},m,m^{\prime},n,n^{\prime},p,p^{\prime}}\int d\widetilde{\boldsymbol{k}}\iiiint dtdt^{\prime}dt^{\prime\prime}dt^{\prime\prime\prime}e^{i\omega t-i\omega_{1}t^{\prime}}e^{-i\omega_{2}t^{\prime\prime}}e^{-i\omega_{3}t^{\prime\prime\prime}}v_{ll^{\prime}}^{a}v_{mm^{\prime}}^{b}v_{nn^{\prime}}^{c}v_{pp^{\prime}}^{d}\\
 &  & \,\,\,\,\,\,\,\,\,\,\,\,\,\times\Theta(t-t^{\prime})\Theta(t^{\prime}-t^{\prime\prime})\Theta(t^{\prime\prime}-t^{\prime\prime\prime})g_{m^{\prime}l}^{a}(t^{\prime}-t)g_{n^{\prime}m}^{a}(t^{\prime\prime}-t^{\prime})g_{p^{\prime}n}^{a}(t^{\prime\prime\prime}-t^{\prime\prime})g_{l^{\prime}p}^{<}(t-t^{\prime\prime\prime})\\
 & = & \frac{\left(-\boldsymbol{e}^{4}\right)}{2\pi\omega_{1}\omega_{2}\omega_{3}}\iint d\widetilde{\boldsymbol{k}}dEf(E)Tr\left(g_{E}^{r-a}v^{d}g_{E_{4}-\hbar\omega_{3}}^{a}v^{c}g_{E-\hbar\omega_{2}-\hbar\omega_{3}}^{a}v^{b}g_{E-\hbar\omega}^{a}v^{a}\right)\\
\\\\\\
\end{eqnarray*}

\begin{eqnarray*}
 &  & \mathscr{\sigma}_{abcd}^{\left(3\right)111}(\omega=\omega_{1}+\omega_{2}+\omega_{3})\\
 & = & \mathscr{\sigma}_{abcd1}^{\left(3\right)111}(\omega=\omega_{1}+\omega_{2}+\omega_{3})+\mathscr{\sigma}_{abcd2}^{\left(3\right)111}(\omega=\omega_{1}+\omega_{2}+\omega_{3})+\mathscr{\sigma}_{abcd3}^{\left(3\right)111}(\omega=\omega_{1}+\omega_{2}+\omega_{3})+\mathscr{\sigma}_{abcd4}^{\left(3\right)111}(\omega=\omega_{1}+\omega_{2}+\omega_{3})\\
 &  & +\mathscr{\sigma}_{abcd5}^{\left(3\right)111}(\omega=\omega_{1}+\omega_{2}+\omega_{3})+\mathscr{\sigma}_{abcd6}^{\left(3\right)111}(\omega=\omega_{1}+\omega_{2}+\omega_{3})+\mathscr{\sigma}_{abcd7}^{\left(3\right)111}(\omega=\omega_{1}+\omega_{2}+\omega_{3})+\mathscr{\sigma}_{abcd8}^{\left(3\right)111}(\omega=\omega_{1}+\omega_{2}+\omega_{3})\\
 & = & \frac{i\left(-\boldsymbol{e}^{4}\right)}{(2\pi)^{2}\omega_{1}\omega_{2}\omega_{3}}\iiint d\widetilde{\boldsymbol{k}}dEdE^{\prime}\left[f(E)-f(E^{\prime})\right]\\
 &  & Tr\left(\frac{v^{a}g_{E+\hbar\omega}^{r}v^{b}g_{E+\hbar\omega_{2}+\hbar\omega_{3}}^{r}v^{c}g_{E^{\prime}}^{r-a}v^{d}g_{E}^{r-a}}{E+\hbar\omega_{3}-E^{\prime}+i\eta}+\frac{g_{E}^{r-a}v^{d}g_{E^{\prime}}^{r-a}v^{c}g_{E-\hbar\omega_{2}-\hbar\omega_{3}}^{a}v^{b}g_{E-\hbar\omega}^{a}v^{a}}{E-\hbar\omega_{3}-E^{\prime}-i\eta}\right)\\
 &  & +\frac{\left(-\boldsymbol{e}^{4}\right)}{(2\pi)^{3}\omega_{1}\omega_{2}\omega_{3}}\iiiint d\widetilde{\boldsymbol{k}}dEdE^{\prime}dE^{\prime\prime}\left[f(E^{\prime})-f(E^{\prime\prime})\right]\\
 &  & Tr\left[\frac{v^{a}g_{E+\hbar\omega}^{r}v^{b}g_{E^{\prime}}^{r-a}v^{d}g_{E^{\prime\prime}}^{r-a}v^{c}g_{E}^{r-a}}{\left(E^{\prime}-\hbar\omega_{3}-E^{\prime\prime}-i\eta\right)\left(E+\hbar\omega_{2}+\hbar\omega_{3}-E^{\prime}+i\eta\right)}+\frac{g_{E}^{r-a}v^{c}g_{E^{\prime\prime}}^{r-a}v^{d}g_{E^{\prime}}^{r-a}v^{b}g_{E-\hbar\omega}^{a}v^{a}}{\left(E^{\prime}+\hbar\omega_{3}-E^{\prime\prime}+i\eta\right)\left(E-\hbar\omega_{2}-\hbar\omega_{3}-E^{\prime}-i\eta\right)}\right]\\
 & = & \left(-\frac{\boldsymbol{e}^{4}}{\hbar^{3}}\right)\frac{i\hbar^{3}}{\omega_{1}\omega_{2}\omega_{3}}\iiint d\widetilde{\boldsymbol{k}}d\widetilde{E}d\widetilde{E}^{\prime}\left[f(E)-f(E^{\prime})\right]\\
 &  & Tr\left(\frac{v^{a}g_{E+\hbar\omega}^{r}v^{b}g_{E+\hbar\omega_{2}+\hbar\omega_{3}}^{r}v^{c}g_{E^{\prime}}^{r-a}v^{d}g_{E}^{r-a}}{E+\hbar\omega_{3}-E^{\prime}+i\eta}+\frac{g_{E}^{r-a}v^{d}g_{E^{\prime}}^{r-a}v^{c}g_{E-\hbar\omega_{2}-\hbar\omega_{3}}^{a}v^{b}g_{E-\hbar\omega}^{a}v^{a}}{E-\hbar\omega_{3}-E^{\prime}-i\eta}\right)\\
 &  & +\left(-\frac{\boldsymbol{e}^{4}}{\hbar^{3}}\right)\frac{\hbar^{3}}{\omega_{1}\omega_{2}\omega_{3}}\iiiint d\widetilde{\boldsymbol{k}}d\widetilde{E}d\widetilde{E}^{\prime}d\widetilde{E}^{\prime\prime}\left[f(E^{\prime})-f(E^{\prime\prime})\right]\\
 &  & Tr\left[\frac{v^{a}g_{E+\hbar\omega}^{r}v^{b}g_{E^{\prime}}^{r-a}v^{d}g_{E^{\prime\prime}}^{r-a}v^{c}g_{E}^{r-a}}{\left(E^{\prime}-\hbar\omega_{3}-E^{\prime\prime}-i\eta\right)\left(E+\hbar\omega_{2}+\hbar\omega_{3}-E^{\prime}+i\eta\right)}+\frac{g_{E}^{r-a}v^{c}g_{E^{\prime\prime}}^{r-a}v^{d}g_{E^{\prime}}^{r-a}v^{b}g_{E-\hbar\omega}^{a}v^{a}}{\left(E^{\prime}+\hbar\omega_{3}-E^{\prime\prime}+i\eta\right)\left(E-\hbar\omega_{2}-\hbar\omega_{3}-E^{\prime}-i\eta\right)}\right]
\end{eqnarray*}

\section{Berry curvature Dipole \label{sec:Ac}}

\begin{eqnarray*}
\bar{\sigma}_{ab}(\omega) & = & \left(-\frac{\boldsymbol{e}^{2}}{\hbar}\right)\frac{\hbar^{2}}{i}\int d\widetilde{\boldsymbol{k}}\sum_{nm}f_{nm}\frac{v_{nm}^{a}v_{mn}^{b}}{\left(\varepsilon_{nk}-\varepsilon_{mk}+i\eta\right)^{2}-\left(\hbar\omega\right)^{2}}\\
 & = & \left(-\frac{\boldsymbol{e}^{2}}{\hbar}\right)\frac{\hbar^{2}}{i}\int d\widetilde{\boldsymbol{k}}\sum_{nm}f\left(\varepsilon_{nk}\right)\frac{v_{nm}^{a}v_{mn}^{b}-v_{mn}^{a}v_{nm}^{b}}{\left(\varepsilon_{nk}-\varepsilon_{mk}+i\eta\right)^{2}-\left(\hbar\omega\right)^{2}}\\
 & = & \left(-\frac{\boldsymbol{e}^{2}}{\hbar}\right)\hbar^{2}\int d\widetilde{\boldsymbol{k}}\sum_{nm}f\left(\varepsilon_{nk}\right)Im\left[\frac{v_{nm}^{a}v_{mn}^{b}-v_{mn}^{a}v_{nm}^{b}}{\left(\varepsilon_{nk}-\varepsilon_{mk}+i\eta\right)^{2}-\left(\hbar\omega\right)^{2}}\right]\\
\Omega_{ab}(k,\omega) & = & \sum_{nm}Im\left[\frac{v_{nm}^{a}v_{mn}^{b}-v_{mn}^{a}v_{nm}^{b}}{\left(\varepsilon_{nk}-\varepsilon_{mk}+i\eta\right)^{2}-\left(\hbar\omega\right)^{2}}\right]\\
 & = & -i\sum_{nm}\frac{v_{nm}^{a}v_{mn}^{b}-v_{mn}^{a}v_{nm}^{b}}{\left(\varepsilon_{nk}-\varepsilon_{mk}+i\eta\right)^{2}-\left(\hbar\omega\right)^{2}}\\
\frac{1}{\hbar}\partial_{c}\Omega_{ab}(k,\omega) & = & -\frac{i}{\hbar}\sum_{nm}\frac{\partial_{c}\left(v_{nm}^{a}\right)v_{mn}^{b}-\partial_{c}\left(v_{mn}^{a}\right)v_{nm}^{b}}{\left(\varepsilon_{nk}-\varepsilon_{mk}+i\eta\right)^{2}-\left(\hbar\omega\right)^{2}}\\
 &  & -\frac{i}{\hbar}\sum_{nm}\frac{v_{nm}^{a}\partial_{c}\left(v_{mn}^{b}\right)-v_{mn}^{a}\partial_{c}\left(v_{nm}^{b}\right)}{\left(\varepsilon_{nk}-\varepsilon_{mk}+i\eta\right)^{2}-\left(\hbar\omega\right)^{2}}\\
\\\frac{1}{\hbar}\partial_{c}\left(v_{nm}^{a}\right) & =\frac{1}{\hbar} & \frac{\partial<n|\frac{1}{\hbar}\frac{\partial H_{k}}{\partial k_{a}}|m>}{\partial k_{c}}=\frac{1}{\hbar}<\partial_{c}n|\frac{1}{\hbar}\frac{\partial H_{k}}{\partial k_{a}}|m>+\frac{1}{\hbar}<n|\frac{1}{\hbar}\frac{\partial H_{k}}{\partial k_{a}}|\partial_{c}m>+v_{nm}^{ac}\\
 & = & v_{nm}^{ac}+\frac{1}{\hbar}\underset{l}{\sum}<\partial_{c}n|l><l|\frac{1}{\hbar}\frac{\partial H_{k}}{\partial k_{a}}|m>+<n|\frac{1}{\hbar}\frac{\partial H_{k}}{\partial k_{a}}|l><l|\partial_{c}m>\\
 & = & v_{nm}^{ac}+\frac{1}{\hbar}\underset{l}{\sum}-<n|\partial_{c}l><l|\frac{1}{\hbar}\frac{\partial H_{k}}{\partial k_{a}}|m>+<n|\frac{1}{\hbar}\frac{\partial H_{k}}{\partial k_{a}}|l><l|\partial_{c}m>\\
 & = & v_{nm}^{ac}+\underset{l}{\sum}<n|\frac{1}{\hbar}\frac{\partial H_{k}}{\partial k_{a}}|l>\frac{<l|\frac{1}{\hbar}\frac{\partial H_{k}}{\partial k_{c}}|m>}{\varepsilon_{m}-\varepsilon_{l}}-\frac{<n|\frac{1}{\hbar}\frac{\partial H_{k}}{\partial k_{c}}|l>}{\varepsilon_{l}-\varepsilon_{n}}<l|\frac{1}{\hbar}\frac{\partial H_{k}}{\partial k_{a}}|m>\\
 & = & v_{nm}^{ac}+\underset{l}{\sum}\frac{v_{nl}^{a}v_{lm}^{c}}{\varepsilon_{m}-\varepsilon_{l}}-\frac{v_{nl}^{c}v_{lm}^{a}}{\varepsilon_{l}-\varepsilon_{n}}\\
\\ &  & D^{abc}\\
 & = & \frac{1}{\hbar}\int dk\partial_{c}\Omega_{ab}(k,\omega)\\
 & = & -i\int dk\underset{mn}{\sum}\frac{v_{nm}^{ac}v_{mn}^{b}-v_{mn}^{ac}v_{nm}^{b}}{\left(\varepsilon_{nk}-\varepsilon_{mk}+i\eta\right)^{2}-\left(\hbar\omega\right)^{2}}\\
 &  & -i\int dk\underset{mn}{\sum}\frac{v_{nm}^{a}v_{mn}^{bc}-v_{mn}^{a}v_{nm}^{bc}}{\left(\varepsilon_{nk}-\varepsilon_{mk}+i\eta\right)^{2}-\left(\hbar\omega\right)^{2}}\\
 &  & -i\int dk\underset{lmn}{\sum}\frac{v_{nl}^{a}v_{lm}^{c}v_{mn}^{b}}{\left[\left(\varepsilon_{nk}-\varepsilon_{mk}+i\eta\right)^{2}-\left(\hbar\omega\right)^{2}\right]\left(\varepsilon_{m}-\varepsilon_{l}\right)}-\frac{v_{nl}^{c}v_{lm}^{a}v_{mn}^{b}}{\left[\left(\varepsilon_{nk}-\varepsilon_{mk}+i\eta\right)^{2}-\left(\hbar\omega\right)^{2}\right]\left(\varepsilon_{l}-\varepsilon_{n}\right)}\\
 &  & +i\int dk\underset{lmn}{\sum}\frac{v_{ml}^{a}v_{ln}^{c}v_{nm}^{b}}{\left[\left(\varepsilon_{nk}-\varepsilon_{mk}+i\eta\right)^{2}-\left(\hbar\omega\right)^{2}\right]\left(\varepsilon_{n}-\varepsilon_{l}\right)}+\frac{v_{ml}^{c}v_{ln}^{a}v_{nm}^{b}}{\left[\left(\varepsilon_{nk}-\varepsilon_{mk}+i\eta\right)^{2}-\left(\hbar\omega\right)^{2}\right]\left(\varepsilon_{l}-\varepsilon_{m}\right)}\\
 &  & -i\int dk\underset{lmn}{\sum}\frac{v_{nm}^{a}v_{nl}^{b}v_{lm}^{c}}{\left[\left(\varepsilon_{nk}-\varepsilon_{mk}+i\eta\right)^{2}-\left(\hbar\omega\right)^{2}\right]\left(\varepsilon_{m}-\varepsilon_{l}\right)}-\frac{v_{mn}^{a}v_{nl}^{c}v_{lm}^{b}}{\left[\left(\varepsilon_{nk}-\varepsilon_{mk}+i\eta\right)^{2}-\left(\hbar\omega\right)^{2}\right]\left(\varepsilon_{l}-\varepsilon_{n}\right)}\\
 &  & +i\int dk\underset{lmn}{\sum}\frac{v_{nm}^{a}v_{ml}^{b}v_{ln}^{c}}{\left[\left(\varepsilon_{nk}-\varepsilon_{mk}+i\eta\right)^{2}-\left(\hbar\omega\right)^{2}\right]\left(\varepsilon_{n}-\varepsilon_{l}\right)}+\frac{v_{nm}^{a}v_{ml}^{c}v_{ln}^{b}}{\left[\left(\varepsilon_{nk}-\varepsilon_{mk}+i\eta\right)^{2}-\left(\hbar\omega\right)^{2}\right]\left(\varepsilon_{l}-\varepsilon_{m}\right)}\\
 & = & -i\int dk\underset{mn}{\sum}\frac{v_{nm}^{ac}v_{mn}^{b}-v_{mn}^{ac}v_{nm}^{b}}{\left(\varepsilon_{nk}-\varepsilon_{mk}+i\eta\right)^{2}-\left(\hbar\omega\right)^{2}}+\frac{v_{nm}^{a}v_{mn}^{bc}-v_{mn}^{a}v_{nm}^{bc}}{\left(\varepsilon_{nk}-\varepsilon_{mk}+i\eta\right)^{2}-\left(\hbar\omega\right)^{2}}\\
 &  & -i\int dk\underset{lmn}{\sum}\frac{v_{nl}^{a}v_{lm}^{c}v_{mn}^{b}-v_{ml}^{c}v_{ln}^{a}v_{nm}^{b}+v_{nm}^{a}v_{nl}^{b}v_{lm}^{c}-v_{nm}^{a}v_{ml}^{c}v_{ln}^{b}}{\left[\left(\varepsilon_{nk}-\varepsilon_{mk}+i\eta\right)^{2}-\left(\hbar\omega\right)^{2}\right]\left(\varepsilon_{m}-\varepsilon_{l}\right)}\\
 &  & -i\int dk\underset{lmn}{\sum}\frac{-v_{nl}^{c}v_{lm}^{a}v_{mn}^{b}+v_{ml}^{a}v_{ln}^{c}v_{nm}^{b}-v_{mn}^{a}v_{nl}^{c}v_{lm}^{b}+v_{nm}^{a}v_{ml}^{b}v_{ln}^{c}}{\left[\left(\varepsilon_{nk}-\varepsilon_{mk}+i\eta\right)^{2}-\left(\hbar\omega\right)^{2}\right]\left(\varepsilon_{l}-\varepsilon_{n}\right)}
\end{eqnarray*}

Comparing the formulas before, we can find that $\bar{\sigma}_{abc,CPGE}^{\left(2\right)}\varpropto\frac{D^{abc}-D^{acb}}{i\omega}$.

 \end{widetext}

\section{Numerical example for disorder effect \label{sec:Ad}}

\begin{figure*}
\includegraphics[width=2\columnwidth]{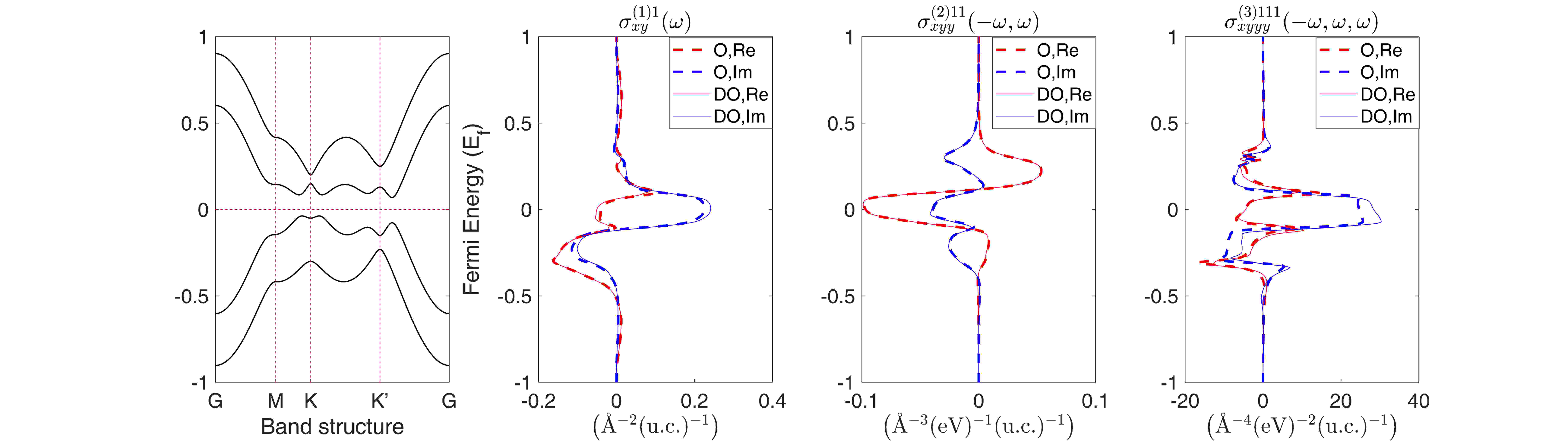} \caption{\textbf{Hexagonal honeycomb graphene model.} (a) The band structure
of the model and first-, second-, third- order optical conductivities.\label{fig:appd}
O,Re (Im) means the real (imagine) part of the conductivies without
disorder and DO means it considers disorder effect.}
\end{figure*}

Generally, Green's function formula is a convenient way to consider
disorder eect in excitation-response process, with some methods like
self energy correction and vertex correction. Here we take hexagonal
honeycomb graphene model as an example, which is a classical model
well-studied for several years for 2D materials properties research.
In the tight-binding approximation, the Hamiltonian could be written
as:

\begin{eqnarray*}
H & = & t\sum_{<ij>\alpha}c_{i\alpha}^{\dagger}c_{j\alpha}+it_{so}\sum_{<<ij>>\alpha\beta}c_{i\alpha}^{\dagger}\nu_{ij}s_{\alpha\beta}^{z}c_{j\beta}\\
 &  & +W\sum_{ij\alpha}c_{i\alpha}^{\dagger}\sigma_{ij}^{z}c_{j\alpha}+M\sum_{i\alpha\beta}c_{i\alpha}^{\dagger}s_{\alpha\beta}^{z}c_{i\beta}\\
 &  & +it_{so}^{R}\sum_{<ij>\alpha\beta}c_{i\alpha}^{\dagger}\left(\boldsymbol{s}\times\boldsymbol{d}_{ij}\right)c_{j\beta}\,,
\end{eqnarray*}
where $t$ is the nearest neighbor hopping term with hopping energy,
and $t_{so}$ is effective intrinsic SOC involving the next-nearest
neighbor hopping. $W$ represents staggered potential breaking inversion
symmetry while $M$ represents exchange eld breaking time reversal
symmetry. The last terms is extrinsic Rashba SOC. In Fig. \ref{fig:appd}(a),
we plot the band structure of the model with the phase of quantum
anomalous Hall effect. The dashed lines in Fig.\ref{fig:appd} (b-d)
are the real part (red) and imaginary part (blue) of response functions
$\mathscr{\sigma}_{xy}^{\left(1\right)1}(\omega)$, $\mathscr{\sigma}_{xyy}^{\left(2\right)11}(-\omega,\omega)$,
$\mathscr{\sigma}_{xyyy}^{\left(3\right)111}(-\omega,\omega,\omega)$
without disorder.

Furthermore, we consider simple disorder $V_{d}=D\sum_{i\alpha}c_{i\alpha}^{\dagger}c_{i\alpha}$,
where D is the disorder strength with 50\% probability to be positive
and negative. Since this kind of disorder just shift the energy up
or down with the same probability, the self-energy correction for
single-particle green function is zero. Besides, we only consider
second-order single-vertex correction (Eq. \ref{eq:v-corr}) for each
$v^{\alpha}$. The response functions with the perturbation of disorder
are plotted with lines in Fig.\ref{fig:appd} (b-d). As we can see,
under weak disorder, the response functions shows a little dierences.\bibliographystyle{unsrt}
\bibliography{Manuscript3.bbl}

\end{document}